\documentclass[a4paper,11pt]{article}
\pdfoutput=1
\usepackage{paperstyle}

\usepackage{amsmath,amssymb,graphicx}
\usepackage{soul}
\usepackage[latin1]{inputenc}
\usepackage{dsfont}
\usepackage{dcolumn}
\usepackage{subfigure}
\usepackage{tabularx}
\usepackage{array}
\newcolumntype{L}{>{\centering\arraybackslash}m{2.6cm}}

\usepackage{mathtools}
\usepackage{bbold}

\usepackage[thinlines]{easytable}

\newcommand{\be}{\begin{eqnarray}}
\newcommand{\en}{\end{eqnarray}}

\newcommand{\As}{A_{\rm{s}}}

\usepackage[section, below, above]{placeins}

\usepackage{afterpage}

\begin{document}

\begin{titlepage}

\vspace*{-15mm}
\vspace*{0.7cm}

\begin{center}

{\Large {\bf What can we learn from the stochastic gravitational wave background produced by oscillons?}}\\[8mm]

Stefan Antusch$^{\star\dagger}$\footnote{Email: \texttt{stefan.antusch@unibas.ch}},  
Francesco Cefal\`{a}$^{\star}$\footnote{Email: \texttt{f.cefala@unibas.ch}} and 
Stefano Orani$^{\star}$\footnote{Email: \texttt{stefano.orani@unibas.ch}}

\end{center}

\vspace*{0.20cm}

\centerline{$^{\star}$ \it
Department of Physics, University of Basel,}
\centerline{\it
Klingelbergstr.\ 82, CH-4056 Basel, Switzerland}

\vspace*{0.4cm}

\centerline{$^{\dagger}$ \it
Max-Planck-Institut f\"ur Physik (Werner-Heisenberg-Institut),}
\centerline{\it
F\"ohringer Ring 6, D-80805 M\"unchen, Germany}

\vspace*{1.2cm}

\begin{abstract}
\noindent The stochastic gravitational wave (GW) background provides a fascinating window to the physics of the very early universe. Beyond the nearly scale-invariant primordial GW spectrum produced during inflation, a spectrum with a much richer structure is typically generated during the preheating phase after inflation (or after some other phase transition at lower energies). This raises the question of what one can learn from a future observation of the stochastic gravitational wave background spectrum about the underlying physics during preheating. Recently, it has been shown that during preheating non-perturbative quasi-stable objects like oscillons can act as strong sources for GW, leading to characteristic features such as distinct peaks in the spectrum. In this paper, we study the GW production from oscillons using semi-analytical techniques. In particular, we discuss how the GW spectrum is affected by the parameters that characterise a given oscillon system, e.g.\ by the background cosmology, the asymmetry of the oscillons and the evolution of the number density of the oscillons. We compare our semi-analytic results with numerical lattice simulations for a hilltop inflation model and a KKLT scenario, which differ strongly in some of these characteristics, and find very good agreement. 
 
\end{abstract}
\end{titlepage}

\tableofcontents

 \newpage
\section{Introduction}
\label{sec:introduction}
The Cosmic Microwave Background (CMB) has proven to be a successful environment to probe the physics of our universe back to the time of recombination but also during inflation \cite{Ade:2015xua,Ade:2015lrj}. Unfortunately, observational constraints on the cosmological history between these two epochs exist only to a limited extent. From the theory point of view inflation must be followed by a process referred to as reheating \cite{Kofman:1994rk,Kofman:1997yn} during which the energy carried by the inflaton is ultimately converted into a thermal bath of radiation via its perturbative decay. In most of the inflationary models, the perturbative decay of the inflaton is preceded by a non-perturbative phase called preheating. The field dynamics during preheating are often very violent and can lead to large field inhomogeneities. This in turn can alter the completion of the reheating process and give rise to the production of a stochastic background of gravitational waves (GWs) \cite{Khlebnikov:1997di,Easther:2006gt,Easther:2006vd,GarciaBellido:2007dg,Dufaux:2007pt,Zhou:2013tsa,Figueroa:2013vif,Ashoorioon:2013oha,Figueroa:2014aya,Figueroa:2016ojl,Antusch:2016con,Figueroa:2017vfa,Liu:2017hua,Antusch:2017flz}.

Different scalar field theories of the early universe lead to qualitatively different preheating dynamics. The latter are (at least to some extent) encoded in the stochastic properties of the resulting background of GWs. Unlike the CMB photons, GWs can propagate almost freely from the time of their production. Hence, they are a promising candidate to fill the observational gap between inflation and the time of recombination. GWs originating from preheating mechanisms are often found to have very high frequencies which are, unfortunately, not accessible to current and planned GW detectors \cite{Abbott:2007kv,TheVirgo:2014hva,AmaroSeoane:2012km,BBO,Kawamura:2011zz,Punturo:2010zz}. However, there are also models predicting GW backgrounds with frequencies within the observable ranges (see e.g\ Refs~\cite{Antusch:2016con,Liu:2017hua,GarciaBellido:2007af,Dufaux:2010cf}).

In this paper we focus on the GWs emitted by so-called oscillons: long-lived, localized, non-linear excitations of real scalar fields. One of the necessary conditions that give rise to the formation of oscillons is a sufficient growth of the initially tiny vacuum fluctuations, as the field oscillates around the minimum of its potential. This can in principle happen via different preheating mechanisms, such as a parametric resonance \cite{Kofman:1994rk,Kofman:1997yn} or tachyonic oscillations \cite{Desroche:2005yt,Brax:2010ai,Antusch:2015nla}. For the fluctuations to form oscillons, the scalar potential must be shallower than quadratic\footnote{An accurate condition has been derived in Ref.~\cite{Amin:2013ika} for small-amplitude oscillons.} in some region around the minimum of the potential.

The formation of oscillons has been studied in various models of inflation \cite{Copeland:2002ku,Broadhead:2005hn,Amin:2011hj,Gleiser:2014ipa,Antusch:2015nla,Lozanov:2017hjm,Hasegawa:2017iay} but also in the context of other field theories \cite{Gleiser:1993pt,Copeland:1995fq,Farhi:2005rz,Fodor:2006zs,Graham:2006vy,Gleiser:2007te,Achilleos:2013zpa}. More recently, in \cite{Antusch:2017flz}, oscillons have been studied in moduli stabilisation scenarios of type IIB string compactification, including the KKLT scenario \cite{Kachru:2003aw}, as well as the case of a discplaced K\"{a}hler modulus in the Large Volume Scenario \cite{Balasubramanian:2005zx,Conlon:2005ki}.

The production of GWs from oscillons has previously been studied in Refs.~\cite{Zhou:2013tsa,Antusch:2016con,Liu:2017hua,Antusch:2017flz} mainly using numerical lattice simulations. In \cite{Zhou:2013tsa} the authors derived analytical estimates for the emitted power in GW from two and four spherically symmetric oscillons and considered also the case of ellipsoidal oscillons. For an axion monodromy model \cite{Silverstein:2008sg,McAllister:2008hb}, they also computed the spectrum of GW using lattice simulations, coming to the conclusion that GW production from oscillons is highly suppressed. On the other hand, in \cite{Antusch:2016con}, oscillons were studied in a potential from hilltop inflation and it was found that a pronounced peak in the GW spectrum can be produced from the oscillons.  

In this paper we investigate the GW production from asymmetric (ellipsoidal) oscillons using a semi-analytical approach which is based on the formalism introduced in \cite{Dufaux:2007pt}. On the basis of simplifying assumptions we derive an analytical expression for the transverse-traceless (TT) part of the energy-momentum tensor (i.e.\ the source of the GWs) for a system of multiple oscillons in an expanding universe. We then use the derived expression to numerically compute the spectrum of GW. We discuss how the parameters that characterise an oscillon system can affect the resulting spectrum of GW and compare our semi-analytical results to those of numerical lattice simulations. We find very good agreement between our semi-analytical approach and the purely numerical lattice simulations.

This paper is structured as follows: In Section~\ref{sec:preh_dynamics}, we discuss the relevant equations for the description of an inhomogeneous scalar field in an expanding universe, as well as the associated production of GWs. In Section~\ref{sec:GW_prod_semi_analytical}, we discuss the production of gravitational waves from oscillons. For a system of multiple, randomly distributed oscillons in an expanding universe we derive an analytical expression for the source of the GWs. In Section~\ref{seq:effects_of_parameters}, we apply our results from Section~\ref{sec:GW_prod_semi_analytical} to study how quantitatively and qualitatively different oscillon systems manifest themselves in the spectrum of GW. In Section~\ref{sec:ex_and_comparison}, we compare results from our semi-analytical approach to those from numerical lattice simulations, before we conclude in Section~\ref{sec:summary_and_conclusions}.

\section{Gravitational waves from an inhomogeneous scalar field in an expanding universe}
\label{sec:preh_dynamics}
In this section we briefly review the production of GWs from an inhomogeneous scalar field in an expanding universe. For more details we refer to Ref.~\cite{Dufaux:2007pt}. Henceforth, we work in natural units $\hbar = c = m_{\rm Pl} = 1/\sqrt{8\pi G} = 1$. However, in most of the cases we write the reduced Planck mass $m_{\rm Pl}$ to explicitly stress the mass dimension.

\subsection{Equations of motion}
\label{sec:preh_EOMs}
We consider a real scalar field $\phi(\textbf{x},t)$ in its potential $V(\phi)$ in a flat Friedman-Lema\^{i}tre-Robertson-Walker (FLRW) background 
\begin{equation}
ds^2 = -dt^2 + a^2(t) \delta_{ij}dx^idx^j\,,
\end{equation}
where $i,j...$ run from 1 to 3.
The dynamics of this system is described by the following set of equations:
\begin{eqnarray}
\ddot{\phi}\, + \, 3H\dot{\phi}\, - \,\frac{1}{a^2}\nabla^2\phi\, + \,\frac{\partial V}{\partial\phi} \, = \, 0\,, \label{eq:EOM_fld} \\
H^2\, = \,\frac{1}{3m^2_{\rm{Pl}}}\left( V\,  + \,\frac{1}{2}\dot{\phi}^2\, + \frac{1}{2a^2}\left|\nabla\phi\right|^2 \right)\,, \label{eq:EOM_hubble}
\end{eqnarray}
where $a$ is the scale-factor and $H = \dot{a}/a$ is the Hubble parameter. In what follows a dot denotes a derivative with respect to cosmic time $t$. Depending on the scalar potential, the dynamics can lead to inhomogeneities of $\phi$, which in turn may give rise to the production of GWs. The latter are represented by the transverse-traceless (TT) part of the metric perturbation, $h_{ij}$, of the FLRW metric (here in the synchronous gauge) 
\begin{equation}
ds^2 = -dt^2 + a^2(t)(\delta_{ij} + h_{ij})dx^idx^j\,.
\end{equation}
The evolution of the gravitational waves is described by 
\begin{equation}
\ddot{h}_{ij}\, + \,3H\dot{h}_{ij}\, - \,\frac{1}{a^2}\nabla^2h_{ij}\, = \,\frac{2}{m^2_{\rm Pl}}\Pi_{ij}^{\rm TT}\,.
\label{eq:EOM_metric_perturbation}
\end{equation}
The tensor $\Pi_{ij}^{\rm TT}$ represents the source of the GWs and corresponds to the TT part of the anisotropic stress 
\begin{equation}
\Pi_{ij} = \frac{1}{a^2}\left(T_{ij} - g_{ij}\langle p\rangle \right)\,,
\label{eq:anisotropic_stress}
\end{equation}
where $\langle p\rangle$ is the homogenous background pressure and $T_{ij}$ denotes the $(i,j)$-th component of the energy-momentum tensor
\begin{equation}
T_{\mu\nu} = \partial_\mu\phi\partial_\nu\phi - g_{\mu\nu}\left(\frac{1}{2}g^{\rho\sigma}\partial_{\rho}\phi\partial_{\sigma}\phi + V \right)\,. 
\label{eq:energy_momentum_tensor}
\end{equation}
After extracting the TT part of eq.~\eqref{eq:anisotropic_stress} and keeping only terms which are first oder in the gravitational coupling, the source term of the GWs reduces to 
\begin{equation}
\Pi_{ij}^{\rm TT} = \frac{1}{a^2}T_{ij}^{\rm TT} =\frac{1}{a^2}\left[\partial_i\phi\partial_j\phi\right]^{\rm TT}\,.
\label{eq:effective_source_term}
\end{equation}
To solve eq.~\eqref{eq:EOM_metric_perturbation} we will work in Fourier space. Adopting the Fourier convention
\be
f(\textbf{x}) = \int \frac{d^3\textbf{k}}{(2\pi)^3}e^{-i\,\textbf{kx}}f(\textbf{k})\,,
\label{eq:Fourier_convention}
\en
eq.~\eqref{eq:EOM_metric_perturbation} in Fourier space reads
\begin{equation}
\ddot{h}_{ij}(\textbf{k},t)\, + \,3H\dot{h}_{ij}(\textbf{k},t)\, + \,\frac{k^2}{a^2}h_{ij}(\textbf{k},t)\, = \,\frac{2}{m^2_{\rm Pl}}\Pi_{ij}^{\rm TT}(\textbf{k},t)\,,
\label{eq:EOM_metric_perturbation_Fourier_space}
\end{equation}
where $k=|\textbf{k}|$ is the magnitude of the \textit{comoving} momentum. The TT part of $\Pi_{ij}$ is
\begin{eqnarray}
\Pi^{\rm TT}_{ij}(\textbf{k},t)\,&=&\,\Lambda_{ij,lm}(\textbf{\^{k}})\,\Pi_{ij}(\textbf{k},t)\nonumber \\
\,&=&\, \left(P_{il}(\textbf{\textbf{\^{k}}})P_{jm}(\textbf{\textbf{\^{k}}}) - \frac{1}{2}\,P_{ij}
(\textbf{\textbf{\^{k}}})P_{lm}(\textbf{\^{k}})\right)\Pi_{ij}(\textbf{k},t) \:,
\label{eq:projection}
\end{eqnarray}
with the tensor $\Lambda_{ij,lm}(\textbf{\^{k}})$ being defined in terms of the projection tensor 
\be
P_{ij}(\textbf{\^{k}}) \equiv \delta_{ij} - \hat{k}_i\hat{k}_j \:,
\en
where $\hat{k}_i \equiv k_i/|\textbf{k}|$. To solve for the GWs we will adopt the formalism discussed in \cite{Dufaux:2007pt}. We therefore rewrite eq.~\eqref{eq:EOM_metric_perturbation_Fourier_space} in terms of conformal time $d\tau=a^{-1}dt$
\be
\bar{h}''_{ij}(\textbf{k},\tau) + \left(k^2 - \frac{a''}{a}\right)\bar{h}_{ij}(\textbf{k},\tau) = \frac{2}{m^2_{\rm Pl}}\,a\,T^{\rm TT}_{ij}(\textbf{k},\tau)\,,
\label{eq:EOM_GW_conformal}
\en
where a prime denotes a derivative with respect to $\tau$. Furthermore, we have defined
\be
\bar{h}_{ij}\equiv a\,h_{ij}\,.
\en
In this paper we are interested in the GWs generated by inhomogeneities of the scalar field on sub-Horizon scales $k^2\gg a^2H^2$. We can therefore drop the term proportional to $a''$ in eq.~\eqref{eq:EOM_GW_conformal}\footnote{For a radiation dominated background we have $a''/a=0$, while for matter domination $a''/a \sim a^2H^2 \ll k^2$.}.
The equations of motion for the GWs then reduce to
\be
\bar{h}''_{ij}(\textbf{k},\tau) + k^2\bar{h}_{ij}(\textbf{k},\tau) = \frac{2}{m^2_{\rm Pl}}\,a\,T^{\rm TT}_{ij}(\textbf{k},\tau)\,.
\label{eq:EOM_GW_conformal_reduced}
\en
Assuming that there is no GW source for $\tau \le \tau_i$, i.e.\ $\bar{h}_{ij}(\textbf{k},\tau)=\bar{h}'_{ij}(\textbf{k},\tau)=0$,  eq.~\ref{eq:EOM_GW_conformal_reduced} has the solution:
\be
\bar{h}_{ij}(\textbf{k},\tau) = \frac{2}{m^2_{\rm Pl}\,k}\int_{\tau_{i}}^\tau d\tau' \sin\left[k(\tau - \tau')\right] \,a(\tau')\,T^{\rm TT}_{ij}(\textbf{k},\tau')\,.
\label{eq:GW_solution}
\en
As long as the source is active, the amplitude of the GWs is given by the solution eq.~\eqref{eq:GW_solution}. Eventually the source vanishes at some moment in time $\tau=\tau_{f}$ and no more GWs are produced. The GWs then freely propagate obeying the equation
\be
\bar{h}''_{ij}(\textbf{k},\tau) + k^2\bar{h}_{ij}(\textbf{k},\tau) =0\,.
\label{eq:EOM_GW_free}
\en
Thus, for $\tau\ge\tau_{f}$ we have
\be
\bar{h}_{ij}(\textbf{k},\tau) =A_{ij}(\textbf{k})\,\sin[k\,(\tau_{f}- \tau)] + B_{ij}(\textbf{k})\,\cos[k\,(\tau_{f}- \tau)].\,
\label{eq:GW_free}
\en
The $\textbf{k}$--dependent coefficients $A_{ij}$ and $B_{ij}$ can be found by requiring that the solutions for $\bar{h}_{ij}(\textbf{k},\tau)$ and $\bar{h}'_{ij}(\textbf{k},\tau)$ from~\eqref{eq:EOM_GW_free} match with the ones from eq.~\eqref{eq:GW_solution} at $\tau=\tau_{f}$. One finds
\begin{align}
B_{ij}(\textbf{k}) &= \bar{h}_{ij}(\textbf{k},\tau_{f}) = \frac{2}{m^2_{\rm Pl}\,k}\int_{\tau_{i}}^{\tau_{f}} d\tau' \sin\left[k(\tau_{f} - \tau')\right] \,a(\tau')\,T^{\rm TT}_{ij}(\textbf{k},\tau')\,,
\label{eq:Bij}
\\
\,A_{ij}(\textbf{k}) &= k^{-1}\,\bar{h}'_{ij}(\textbf{k},\tau_{f}) = \frac{2}{m^2_{\rm Pl}\,k}\int_{\tau_{i}}^{\tau_{f}} d\tau' \cos\left[k(\tau_{f} - \tau')\right] \,a(\tau')\,T^{\rm TT}_{ij}(\textbf{k},\tau')\,.
\label{eq:Aij}
\end{align}

\subsection{Energy density and spectrum of GW}
\label{sec:Edensity_and_spectrum}
The energy density of a GW cannot be localized within a wavelength but is instead defined as an average over a portion of (\textit{comoving}) volume $\mathcal{\mathcal{V}}$ that contains several wavelengths of the emitted GWs:
\begin{align}
\rho_{\rm GW} &= \frac{m^2_{\rm Pl}}{4} \langle \dot{h}_{ij}(\textbf{x},t)\dot{h}_{ij}(\textbf{x},t) \rangle_{\mathcal{V}} \,
\label{eq:rhoGW}\\
&\simeq \frac{m^2_{\rm Pl}}{4a^4} \langle \bar{h}'_{ij}(\textbf{x},\tau)\bar{h}'_{ij}(\textbf{x},\tau)\rangle_{\mathcal{\mathcal{V}}}\,
\label{eq:rhoGW_conformal}\\
&= \frac{m^2_{\rm Pl}}{4a^4} \frac{1}{\mathcal{V}}\int_\mathcal{\mathcal{V}} d^3{\textbf x} \bar{h}'_{ij}(\textbf{x},\tau)\bar{h}'_{ij}(\textbf{x},\tau)\, 
\\
&= \frac{m^2_{\rm Pl}}{4a^4} \frac{1}{\mathcal{V}}\int \frac{d^3\textbf{k}d^3\textbf{k}'}{(2\,\pi)^6} \int_\mathcal{\mathcal{V}} d^3\textbf{x}\,e^{i (\textbf{k}+\textbf{k}')\textbf{x}}\,\bar{h}'_{ij}(\textbf{k},\tau)\bar{h}'_{ij}(\textbf{k}',\tau)\,\\
&= \frac{m^2_{\rm Pl}}{4a^4} \frac{1}{\mathcal{V}}\int \frac{d^3\textbf{k}}{(2\,\pi)^3}\,\bar{h}'_{ij}(\textbf{k},\tau)\bar{h}'^*_{ij}(\textbf{k},\tau)\,
\end{align}
where in eq.~\eqref{eq:rhoGW_conformal} we dropped two terms propotional to $a\,H$ and $a^2\,H^2$, respectively, since they are negligible when considering sub-Horizon wavelengths. We can now write down the spectrum of GWs per logarithmic momentum interval
\be
\Omega_{\rm GW}(k,\tau)=\frac{1}{\rho_{\rm c}}\frac{d\rho_{\rm GW}}{d\ln\,k}=\frac{m^2_{\rm Pl}\,k^3}{\rho_{\rm c}\,4a^4} \frac{1}{\mathcal{V}}\int \frac{d\Omega}{(2\,\pi)^3}\,\bar{h}'_{ij}(\textbf{k},\tau)\bar{h}'^*_{ij}(\textbf{k},\tau)\,,
\label{eq:GW_spec_normalized}
\en
where $d\Omega$ is a solid angle in momentum space, $k$ is the magnitude of the comoving momentum and. The energy spectrum of the GWs is as usual rescaled by the critical density of the universe $\rho_{\rm c}$. We can now plug the solution~\eqref{eq:GW_free} into eq.~\eqref{eq:GW_spec_normalized}
\begin{align}
\Omega_{\rm GW}(k,\tau)&=\frac{m^2_{\rm Pl}\,k^3}{\rho_{\rm c}\,4a^4} \frac{1}{\mathcal{\mathcal{V}}}\int \frac{d\Omega}{(2\,\pi)^3}\,\langle\bar{h}'_{ij}(\textbf{k},\tau)\bar{h}'^*_{ij}(\textbf{k},\tau)\rangle_{T}  \nonumber\\
&=\frac{m^2_{\rm Pl}\,k^3}{\rho_{\rm c}\,4a^4} \frac{1}{\mathcal{\mathcal{V}}}\int \frac{d\Omega}{(2\,\pi)^3}\frac{k^2}{2}\sum_{i,j}(|A_{ij}(\textbf{k})|^2+|B_{ij}(\textbf{k})|^2)\,,
\label{eq:GW_spec_normalized}
\end{align}
where the GWs with momentum $k$ have been averaged over one period of oscillation $T=2\pi/k$. With the expressions~\eqref{eq:Bij} and~\eqref{eq:Aij} for $B_{ij}$ and $A_{ij}$ respectively, we can bring the spectrum $\Omega_{\rm GW}(k,\tau)$ to the following final form \cite{Dufaux:2007pt}
\begin{align}
\Omega_{\rm GW}(k,\tau) =& \;
\frac{k^3}{2\,a^4\,\rho_{\rm c}\,m^2_{\rm Pl}}\frac{1}{\mathcal{\mathcal{V}}}\int \frac{d\Omega}{(2\,\pi)^3}\sum_{i,j}\left[\left| \int_{\tau_{i}}^{\tau_{f}} d\tau' \cos(k\tau') \,a(\tau')\,T^{\rm TT}_{ij}(\textbf{k},\tau') \right|^2 \right. \nonumber \\
&+\left. \left| \int_{\tau_{i}}^{\tau_{f}} d\tau' \sin(k\tau') \,a(\tau')\,T^{\rm TT}_{ij}(\textbf{k},\tau') \right|^2 \right] .
\label{eq:omega_k_tau}
\end{align}
In the next section we describe our method to calculate the spectrum~\eqref{eq:omega_k_tau} produced by asymmetric oscillons.

\section{Gravitational radiation from oscillons}
\label{sec:GW_prod_semi_analytical}

Oscillons can vary in shape and size as they evolve with time. The non-linear dynamics governing their evolution gives rise to interactions with their environment and in particular among themselves. Such interactions can lead to temporary large and irregular deformations of the spatial oscillon profile and to enhanced (or reduced) field amplitudes. However, if oscillons are sufficiently separated from each other, lattice simulations have shown that both their maximum amplitude and their physical size can be approximately constant over many oscillations. 

Lattice simulations have also shown that oscillons can be an active source of gravitational wave production, unless they are spherically symmetric. In this section, we want to get an analytical understanding of the GW emission from asymmetric oscillons. The ultimate goal is to obtain an analytic expression for the spectrum of GW produced by asymmetric oscillons, that can be evaluated numerically. 

\begin{figure}[ht]
\centering
\includegraphics[width=0.9\textwidth]{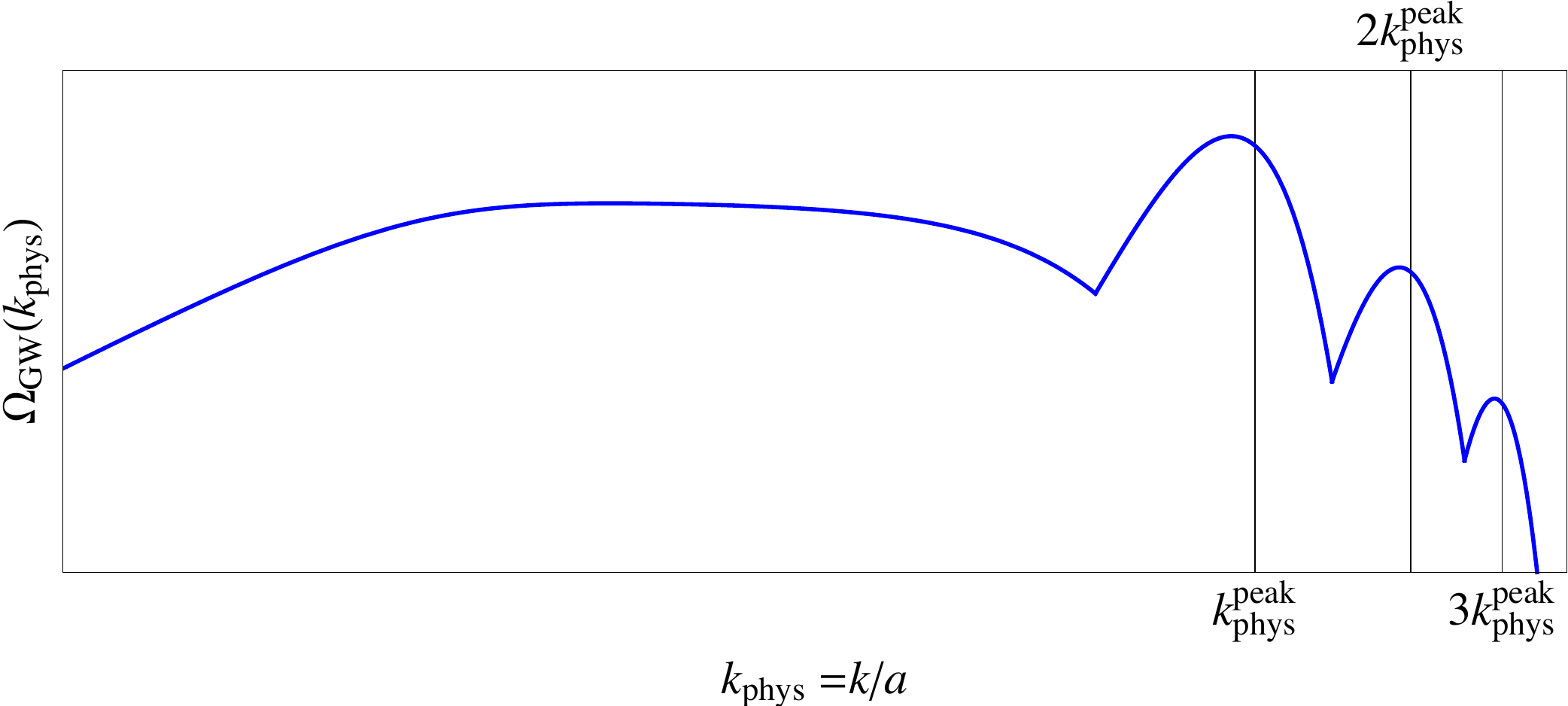}
\caption{Schematic plot of a typical spectrum of GW from oscillons. The spectrum features a multiple peak structure with a dominant peak at a characteristic \textit{physical} wavenumber $k^{\rm peak}_{\rm phys}$ which corresponds to the oscillation frequency of the source (i.e.\ the frequency of the square of the time dependent oscillon amplitude). Other peaks appear at multiples of $k^{\rm peak}_{\rm phys}$ and correspond to the higher harmonics of the source. Moreover, the spectrum typically features a plateau (or broader peak) for $k_{\rm phys} \ll k^{\rm peak}_{\rm phys}$. The latter is not necessarily related to the oscillon dynamics and can also originate from the dynamics preceding the formation of oscillons.} 
\label{fig:schematic_GW}
\end{figure}

Figure~\ref{fig:schematic_GW} shows a schematic illustration of a typical stochastic background of GW produced in the presence of oscillons (see e.g.\ \cite{Zhou:2013tsa,Antusch:2016con,Antusch:2017flz}) as a function of the \textit{physical} wavenumber
\be
k_{\rm phys} = k/a\,.
\en

As illustrated in Figure~\ref{fig:schematic_GW}, the spectrum has a peak structure which typically consists of multiple distinct peaks with a dominant peak at a characteristic \textit{physical} wavenumber $k^{\rm peak}_{\rm phys}$ and further additional peaks at multiples of $k^{\rm peak}_{\rm phys}$ which are, however, less and less pronounced towards the UV. This structure originates from the oscillon dynamics with the positions of the peaks corresponding essentially to the oscillation frequency of the source and its higher harmonics. Below the characteristic scale $k^{\rm peak}_{\rm phys}$, the spectrum shows a relatively flat plateau that typically falls off towards the IR. Depending on model specific dynamics, the plateau can be more ore less pronounced and may not even be directly related to the oscillon dynamics, but rather to an earlier phase of preheating. In what follows, we will concentrate on reproducing the dominant part of the spectrum that is generated from the dynamics of asymmetric oscillons. 
In order to describe the oscillons and their dynamics we will make the following three simplifying assumptions:

\begin{enumerate}
\item For the field profile of an oscillon we assume a three dimensional Gaussian with constant \textit{physical} width (see below). In comoving coordinates we can therefore write the field configuration of a single oscillon at a position $\textbf{x}^0=(x^0,y^0,z^0)^T$ as
\be
\phi_{\rm oscillon}(\textbf{x},t) = \Phi(t)\mathcal{F}(\textbf{x},t)
\en
where $\Phi(t)$ is the time dependent amplitude of the oscillon and $\mathcal{F}(\textbf{x},t)$ is the spatial profile, given by
\be
\mathcal{F}(\textbf{x},t) = e^{\frac{-a^2(t)}{2}\frac{(\textbf{x}-\textbf{x}^0)^2}{\textbf{R}^2}} = e^{\frac{-a^2(t)}{2}\frac{(x-x^0)^2}{R_x^2}+\frac{(y-y^0)^2}{R_y^2}+\frac{(z-z^0)^2}{R_z^2}}\,.
\label{eq:oscillon_profile}
\en
The \textit{comoving} widths of the oscillon in the $x$, $y$ and $z$ direction are expressed in terms of the (constant) \textit{physical} widths $R_x$, $R_y$ and $R_z$. Note, that the time dependence in the spatial profile $\mathcal{F}$ is only due to the fact that the \textit{comoving} width of an oscillon will be time dependent in an expanding universe, when assuming the \textit{physical} width to be constant. 
\item To quantify the asymmetry of an oscillon we assume that one of the widths, e.g.\  $R_y$, is different from the others. We can therefore eliminate one parameter and write 
\be
R_x=R_z\equiv R,\quad \text{and}\quad R_y=R\,(1+\Delta)\,, 
\en
where $\Delta$ parametrizes the asymmetry of the oscillon. In general we have 
\be
\mathcal{O}(0.1) \lesssim \Delta \lesssim \mathcal{O}(1)\,.
\en
\end{enumerate}

\subsection{The GW source for a single oscillon}
\label{sec:single_osc}
We begin our discussion with the case of a single oscillon sitting at the origin $\textbf{x}^0=(0,0,0)^T$.
The case of multiple oscillons is discussed later in Section~\ref{sec:multiple_osc}.
Based on the assumptions above we can compute the TT part of the energy-momentum tensor $T^{\rm TT}_{ij}$ of a single oscillon from eq.~\eqref{eq:projection}
\be
T^{\rm TT}_{ij}(\textbf{k},t)&=&\Lambda_{ij,lm}(\textbf{\^{k}})\,T_{lm}(\textbf{k},t) = \Phi^2(t)\,\Lambda_{ij,lm}(\textbf{\^{k}})\mathcal{T}_{lm}(\textbf{k},t)\,,
\en
with $\mathcal{T}_{ij}(\textbf{k},t)\equiv \Phi^{-2}(t)T_{ij}(\textbf{k},t)$ given by
\begin{eqnarray}
\mathcal{T}_{ij}(\textbf{k},t)&=&\int d^3\textbf{x}\,e^{-i\textbf{k}\textbf{x}}\partial_i\mathcal{F}(\textbf{x},t)\partial_j\mathcal{F}(\textbf{x},t)\, \nonumber\\
&=&a^2(t) \int d^3\textbf{x}\,e^{-i\textbf{k}\textbf{x}}\,e^{-a^2(t)\,x_s^2R^{-2}_s}\frac{x_i\,x_j}{R^2_i\,R^2_j}\, \nonumber\\
&=& -\frac{\pi^{3/2}}{4\,a^3(t)}\,e^{-\frac{k^2_x\,R_x^2+k^2_y\,R_y^2+k^2_z\,R_z^2}{4\,a^2(t)}}R_x\,R_y\,R_z\,\left(k_i k_j - \delta_{ij}\frac{2\,a^2(t)}{R_i\,R_j}\right)\,\nonumber\\
&=& \,e^{-\frac{R^2\left(k_x^2+k_z^2+k_y^2(1+\Delta)^2\right)}{4\,a^2(t)}}\,\mathcal{S}_{ij}\,,
\end{eqnarray}
where we have defined
\be
\mathcal{S}_{ij} \equiv -\frac{\pi^{3/2} R^3\,(1 + \Delta)}{4\,a^3(t)}\left[k_i k_j - \frac{2\,a^2(t)}{R^2}\left(\delta_{ij} + \delta_{is}\delta_{jr}\frac{\delta_{s2}\delta_{r2}\Delta(2+\Delta)}{(1+\Delta)^2}\right) \right]\,,
\en
in the last equality. Furthermore we have chosen the direction of the asymmetry to be the $y$-direction (or ``2-direction''), i.e.\
\be
R_x=R_z \equiv R,\quad \text{and}\quad R_y=R\,(1+\Delta)\,.
\en
For the TT part of $\mathcal{T}_{ij}(\textbf{k},t)$ we find
\begin{eqnarray}
\mathcal{T}^{TT}_{ij}(\textbf{k},t)&=&\Lambda_{ij,lm}(\textbf{\^{k}})\mathcal{T}_{lm}(\textbf{k},t)\,\nonumber\\
&=&e^{-\frac{R^2\left(k_x^2+k_z^2+k_y^2(1+\Delta)^2\right)}{4\,a^2(t)}}\,\frac{\pi ^{3/2} \Delta  (\Delta +2) R}{4 a(t) (\Delta +1)}\,f_{ij}(\textbf{k})\,,
\label{eq:TT_part_fancyTij}
\end{eqnarray}
with $f_{ij}(\textbf{k})$ given by
\be
\left(f_{ij}\right)=\frac{1}{|\textbf{k}|^4}\, \begin{pmatrix}
-k_x^2 k_y^2 + (k_x^2 + k_y^2) k_z^2 + k_z^4 & k_x k_y (k_x^2 + k_z^2) & -k_x k_z (k_x^2 + 2 k_y^2 + k_z^2)  \\
k_x k_y (k_x^2 + k_z^2) & -(k_x^2 + k_z^2)^2 & k_y k_z (k_x^2 + k_z^2)  \\
-k_x k_z (k_x^2 + 2 k_y^2 + k_z^2) & k_y k_z (k_x^2 + k_z^2)  & k_x^4 - k_y^2 k_z^2 + k_x^2 (k_y^2 + k_z^2)  
\end{pmatrix} \!\!.
\en
Note, that from eq.~\eqref{eq:TT_part_fancyTij} it is immediately evident that spherically symmetric oscillons (i.e.\ $\Delta=0$) will not emit gravitational waves, since
\be
\mathcal{T}^{TT}_{ij}(\textbf{k},t) = 0\,,\qquad\textrm{for}\qquad \Delta=0\,.
\en 

\subsection{The GW source for multiplte oscillons}
\label{sec:multiple_osc}
Let us now discuss the generalized case of multiple oscillons. 
In the case of $N$ oscillons with identical profile but different positions $\textbf{x}^q$ we can write the field as
\be
\phi_{\rm multi}(\textbf{x},t)=\sum_{q=1}^N \Phi_q(t) e^{\frac{-a^2(t)}{2}\frac{(\textbf{x}-\textbf{x}^q)^2}{\textbf{R}^2}} \equiv \sum_{q=1}^N \Phi_q(t)\,\mathcal{F}_q(\textbf{x},t) \,,
\en
with
\be
\mathcal{F}_q(\textbf{x},t) \equiv e^{\frac{-a^2(t)}{2}\frac{(\textbf{x}-\textbf{x}^q)^2}{\textbf{R}^2}}\,.
\en
The function $\Phi_q(t)$ describes the time dependent amplitude of the oscillon at position $\textbf{x}^q$. In general, we assume that all the $\Phi_q$ have different phases but same periodicity and the same amplitude. The TT part of the energy-momentum tensor of this system is
\begin{eqnarray}
T^{\rm TT,\,\rm multi}_{ij}&=&\left[\partial_i\phi_{\rm multi}\partial_j\phi_{\rm multi}\right]^{\rm TT}\nonumber\\
&=&\left[\partial_i\left(\sum_{q} \Phi_q(t)\,\mathcal{F}_q(\textbf{x},t)
\right) \partial_j \left(\sum_{q} \Phi_q(t)\,\mathcal{F}_q(\textbf{x},t)
\right)\right]^{\rm TT}\nonumber\\
&=&\left[\sum_{q} \Phi^2_q(t)\,\partial_i\mathcal{F}_q(\textbf{x},t)\partial_j\mathcal{F}_q(\textbf{x},t)+ \sum_{q\neq r} \Phi_q(t)\Phi_r(t)\,\partial_i\mathcal{F}_q(\textbf{x},t)\partial_j\mathcal{F}_r(\textbf{x},t)
 \right]^{\rm TT}\,.
 \label{eq:TijTT_multi_with_interference}
\end{eqnarray}
The second sum accounts for contributions from the interference between two oscillons at different positions $\textbf{x}^q$ and $\textbf{x}^r$. These contributions are exponentially suppressed with the square of their distance. We will assume that the oscillons are sufficiently separated from each other and neglect the interference term, i.e.\ 
\be
T^{\rm TT,\,\rm multi}_{ij}\simeq\left[\sum_{q} \Phi^2_q(t)\,\partial_i\mathcal{F}_q(\textbf{x},t)\partial_j\mathcal{F}_q(\textbf{x},t)\right]^{\rm TT}\,,
\en
which in Fourier space reads
\begin{eqnarray}
T^{\rm TT,\,\rm multi}_{ij}(\textbf{k},t)&\simeq&\Lambda_{ij,lm}(\textbf{\^{k}})\sum_{q} \Phi^2_q(t)\, \int d^3\textbf{x}\,e^{-i\textbf{k}\textbf{x}}\partial_l\mathcal{F}_q(\textbf{x},t)\partial_m\mathcal{F}_q(\textbf{x},t)\nonumber\\
&=&\Lambda_{ij,lm}(\textbf{\^{k}})\,\mathcal{T}_{lm}(\textbf{k},t)\sum_{q} \Phi^2_q(t)\,e^{-i\,\textbf{k}\,\textbf{x}^q}\,.
\label{eq:TijTT_multi}
\end{eqnarray}
Once we have specified all free parameters, as well as the background cosmology, we can compute the spectrum of the GW produced by a given system of $N$ oscillons   distributed throughout a given \textit{comoving} volume $\mathcal{\mathcal{V}}$ according to eq.~\eqref{eq:omega_k_tau}:
\begin{align}
\Omega_{\rm GW}(k,\tau)&=
\frac{k^3}{2\,a^4\,\rho_{\rm c}\,m^2_{\rm Pl}}\frac{1}{\mathcal{\mathcal{V}}}\int \frac{d\Omega}{(2\,\pi)^3}\sum_{i,j}\left[\left| \int_{\tau_{i}}^{\tau_{f}} d\tau' \cos(k\tau') \,a(\tau')\,T^{\rm TT,\,\rm multi}_{ij}(\textbf{k},\tau') \right|^2\right.\nonumber\\
&+ \left.\left| \int_{\tau_{i}}^{\tau_{f}} d\tau' \sin(k\tau') \,a(\tau')\,T^{\rm TT,\,\rm multi}_{ij}(\textbf{k},\tau') \right|^2 \right]\,.
\label{eq:omega_k_tau_multi}
\end{align}
To evaluate eq.~\eqref{eq:omega_k_tau_multi} we transform to spherical coordinates in $k$-space and approximate the integral over the solid angle by a discrete sum as
\be
\int d\Omega \rightarrow \sum_{i,j = 0}^{N_{\rm angles}} \sin\theta_i \,\Delta\theta\,\Delta\phi\,,
\en
where
\be
\theta_j = j\, \Delta\theta = j\,\frac{\pi}{N_{\rm angles}},\qquad\textrm{and}\qquad\phi_j = j\, \Delta\phi = j\,\frac{2\pi}{N_{\rm angles}}\,.
\en
The sum over the components $i,j$ of the stress-energy tensor and the integral over $d\tau'$ is then performed for fixed values of $\theta_j$, $\phi_j$ and $k$, where the integral over $d\tau'$ is computed numerically.

In the following section we will discuss the possible effects of parameters characterising an oscillon system on the resulting spectrum of GWs.

\section{Effects of different parameters on $\Omega_{\rm GW}$}
\label{seq:effects_of_parameters}
The formation and evolution of oscillons are model dependent processes that are ultimately related to the shape of the scalar potential, particularly around the minimum. Different models can a priori lead to qualitatively and quantitatively different field dynamics that may be directly reflected in the characteristics of the resulting background of GW. This implies that a direct observation of such a stochastic signal could in principle be used to gain valuable information about the underlying theory. It is thus crucial to understand how model specific features, as e.g.\ the background cosmology or the oscillon dynamics itself, are imprinted in the spectrum of GWs.

In this section we discuss how the physical properties of a given system of $N$ oscillons at different positions $\textbf{x}^q$ manifest themselves in the spectrum of GWs. To this end we numerically evaluate eq.~\eqref{eq:omega_k_tau_multi} as described above for different realisations of an $N$ oscillon system and compare the obtained results. Henceforth, all quantities denoted by a subscript ``0'' will indicate the value of the corresponding quantity at the time when oscillons become dynamically relevant.

So far we did not specify the function $\Phi_q(t)$ describing the time-dependent amplitude of an oscillon at position $\textbf{x}^q$. 
For the purpose of our discussion we assume that all $N$ oscillons have the same (constant) maximum amplitude and the same period of oscillation. Explicitly, we assume
\be
\Phi_q(t) = A\,\cos(\omega_{\rm osc}t +\varphi_q)\,,
\label{eq:oscillon_amplitude}
\en
where $\omega_{\rm osc}$ is the oscillation frequency of the oscillons, $A$ is the maximum amplitude, i.e.\ essentially the amplitude at the center of the oscillons and $\varphi_q$ is a random phase. As mentioned in Section~\ref{sec:GW_prod_semi_analytical}, the spectrum of GWs exhibits a characteristic peak structure, where the positions of the peaks in (\textit{physical}) $k$-space correspond to the main frequency of the source (and its higher harmonics) (cf.\ Figure~\ref{fig:schematic_GW}). Thus, assuming a simple cosine will result in a single peak in the GW spectrum. Moreover, note that the source is proportional to the square of the amplitude ($T^{\rm TT}_{ij}\propto \Phi^2_q$) (see\ eq.~\eqref{eq:TijTT_multi}). When assuming a cosine for $\Phi_q(t)$, the position of the peak in the GW spectrum will therefore correspond to twice the oscillation frequency of the oscillon (i.e.\ $k^{\rm peak}_{\rm phys} = 2\omega_{\rm osc}$). In general, this is not necessarily the case. For asymmetric potentials, the dominant oscillation frequency of the oscillon and the position of the peak are identical (since the square of the amplitude does not have a different periodicity).

Eq.~\eqref{eq:oscillon_amplitude} is exactly the solution to the equation of motion of a homogeneous scalar field with a quadratic potential 
\be
V(\Phi) = \frac{m^2}{2}\Phi^2\,,
\en
with $m=\omega_{\rm osc}$. We use this fact to set the initial energy density when evaluating eq.~\eqref{eq:omega_k_tau_multi}. On the basis of our experience (i.e.\ from numerical lattice simulations) we assume that the average energy density $\rho_0$ at the time when oscillons become relevant is much smaller than the maximal potential energy at the center of an oscillons, i.e.\
\be
\rho_0 = 3\,H^2_0 \ll \frac{V(A)}{3}\,.
\en
We also assume that the size of the comoving volume $\mathcal{\mathcal{V}}$ is initially (at $a=a_0=1$) much larger than the volume covered by a sphere with a radius corresponding to the width of the oscillons $R$:
\be
\mathcal{\mathcal{V}}\gg R^3\,.
\en
In cases with multiple oscillons we assume that the minimum distance between two oscillons $d_{\rm min}$ is not smaller than four times the width of an oscillon
\be
d_{\rm min} \ge 4R\,.
\en
In the following part of this section we are going to discuss the effects of various parameters on the spectrum of GWs. To this end we sometimes explicitly compute the spectrum $\Omega_{\rm GW}(k)$ according to eq.~\eqref{eq:omega_k_tau_multi}. If not otherwise stated, we fix the amplitude of the oscillons $A$, the width of the oscillons $R$, the oscillation frequency $\omega_{\rm osc}$, the initial Hubble parameter $H_0$ and the comoving volume $\mathcal{\mathcal{V}}$ to the following (typical) values:
\be
A=0.05\,m_{\rm Pl}\,,\qquad \omega_{\rm osc}/m_{\rm Pl}=m_{\rm Pl} R=1\,,
\en
\be
H_0 = \frac{1}{10}\sqrt{\frac{\frac{\omega^2_{\rm osc}}{2}A^2}{3\,m^2_{\rm Pl}}} \simeq 0.002 m_{\rm Pl}\,, \quad \textrm{and} \quad \mathcal{V} = (100\,R)^3\sim\mathcal{O}\left(\frac{H^{-3}_0}{100}\right)\,.
\en
For the background cosmology we use a matter dominated background, with
\be
a(t) = \left(1+\frac{3 H_0 t}{2}\right)^{2/3}
\en
such that 
\be
a(0) = 1,\quad\textrm{and} \quad \left.\frac{\dot{a}(t)}{a(t)}\right|_{t=0} = H_0\,.
\en

\begin{figure}[ht]
\centering
\includegraphics[width=0.9\textwidth]{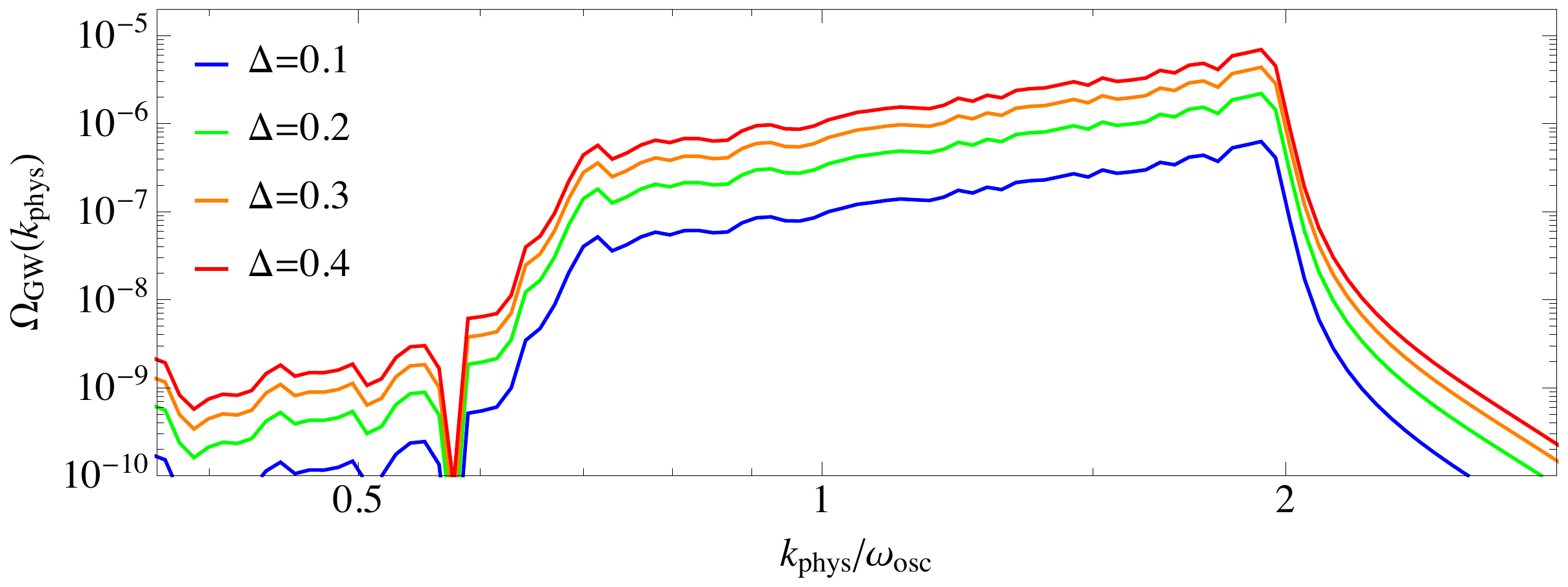}
\caption{The spectrum $\Omega_{\rm GW}$ at $a(\tau_f)\equiv a_f = 3$ as a function of the physical wavenumber $k_{\rm phys}$ for a single oscillon in a matter dominated background. The spectrum is shown for $\Delta=0.1$ (blue), $\Delta=0.2$ (green), $\Delta=0.3$ (orange) and $\Delta=0.4$ (red).} 
\label{fig:spectra_ratios_differentDelta}
\end{figure}
\subsection{The amplitude $A$ of the oscillons}
We begin our discussion with the effect of the amplitude $A$ of the oscillons which does not even require the evaluation of eq.~\eqref{eq:omega_k_tau_multi} explicitly. Indeed, by replacing the $\Phi_q$ with eq.~\eqref{eq:oscillon_amplitude} in the expression for $T^{\rm TT,\,\rm multi}_{ij}$ eq.~\eqref{eq:TijTT_multi}, we can factor out the amplitude $A$ in the expression for $\Omega_{\rm GW}$ eq.~\eqref{eq:omega_k_tau_multi} and find
\be
\Omega_{\rm GW} \propto A^4\,.
\en
This means that a change in the amplitude $A$ results simply in an overall rescaling of the spectrum of GWs. Nevertheless, since $\Omega_{\rm GW}$ scales with the fourth power of the amplitude, small changes in $A$ can have a significant impact on the magnitude of the spectrum. 

\subsection{The impact of the asymmetry $\Delta$}
In order to get a first understanding of the effect of $\Delta$ on the spectrum of GW recall eq.~\eqref{eq:TT_part_fancyTij} where for an asymmetry in the $y$-direction we find that
\be
\mathcal{T}^{TT}_{ij} \propto e^{-\frac{R^2\,k_y^2(2\Delta + \Delta^2)}{4\,a^2(\tau)}}\,\frac{\Delta  (\Delta +2)}{(\Delta +1)}\,.
\label{eq:delta_dependence}
\en
Together with eq.~\eqref{eq:omega_k_tau_multi} this tells us that the spectrum of GWs (for small $\Delta$) scales as
\be
\Omega_{\rm GW} \propto \Delta^2 + {\cal O}(\Delta^4)\,,
\label{eq:omega_scaling_delta}
\en
with a potential small time-dependence of the ${\cal O}(\Delta^4)$-term from the exponential in eq.~\eqref{eq:delta_dependence}. The value of the exponential is of course model dependent in the sense that it depends on the width of the oscillons, as well as on the magnitude $k$ of the relevant wave numbers. 

Figure~\ref{fig:spectra_ratios_differentDelta} shows the spectrum $\Omega_{\rm GW}(k_{\rm phys})$ as a function of the physical wavenumber $k_{\rm phys} = a^{-1}k$ calculated from eq.~\eqref{eq:omega_k_tau_multi} for a single oscillon ($N=1$) in a matter dominated background  at $a(\tau_f)\equiv a_f =3$. The spectrum is shown for different values of $\Delta$, illustrating that the shape of the spectrum remains practically unaffected by changing the asymmetry $\Delta$: increasing $\Delta$ simply leads to an overall increase in $\Omega_{\rm GW}$.

\subsection{The background cosmology}
Both effects considered so far led to a simple overall rescaling of $\Omega_{\rm GW}$. The situation is different when considering different background cosmologies, or in other words a different equation of state parameter $w$. In fact, we find that changing $w$ does not only lead to an overall rescaling of the spectrum, but also changes its slope (on a $\log$-$\log$ scale). 

\begin{figure}[ht]
\centering
\includegraphics[width=\textwidth]{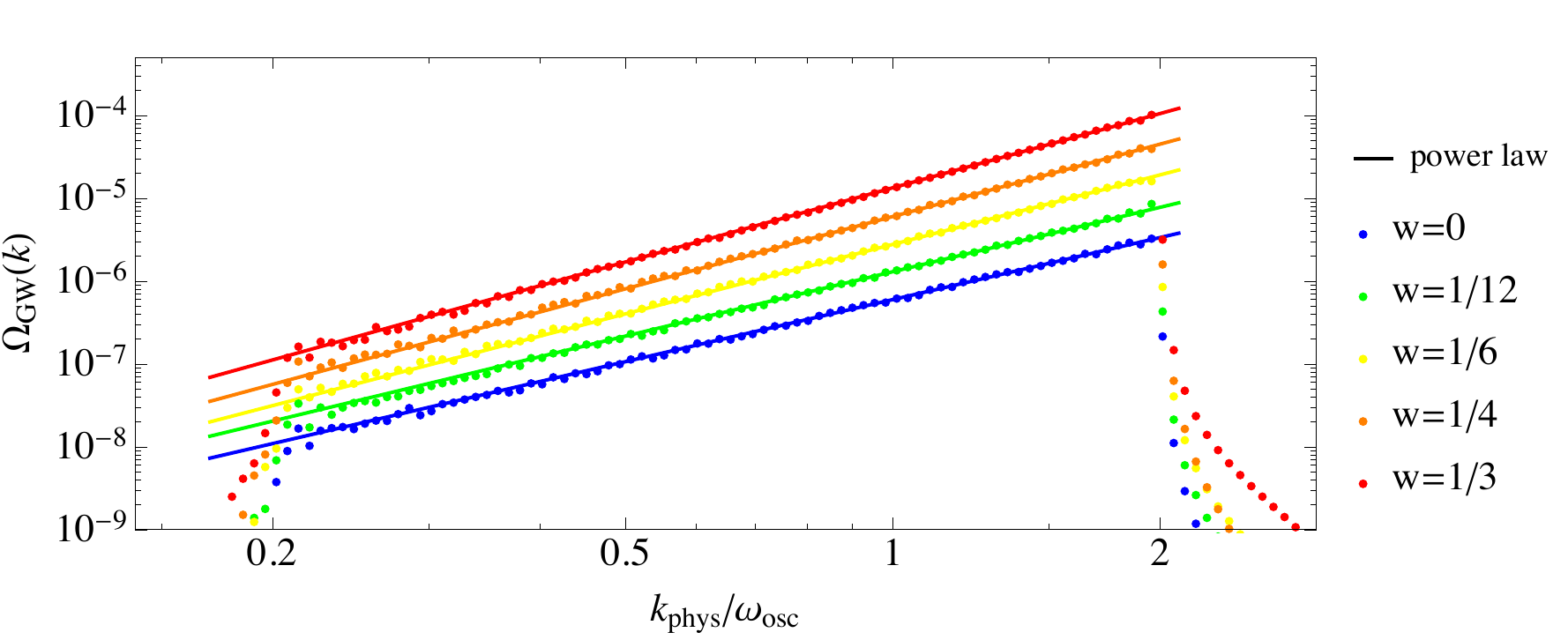}
\caption{The spectrum $\Omega_{\rm GW}(k)$ computed at $a(\tau_f)\equiv a_f=10$ for different equation of state parameters: $w=0$ (blue), $w=1/12$ (green), $w=1/6$ (yellow), $w=1/4$ (orange), and $w=1/3$ (red). One can clearly see that not only the amplitude of the spectrum increases when increasing $w$ but also the slope on a log-log scale.} 
\label{fig:spectra_different_w}
\end{figure}

\begin{figure}[ht]
\centering
\subfigure{\includegraphics[width=0.505\textwidth]{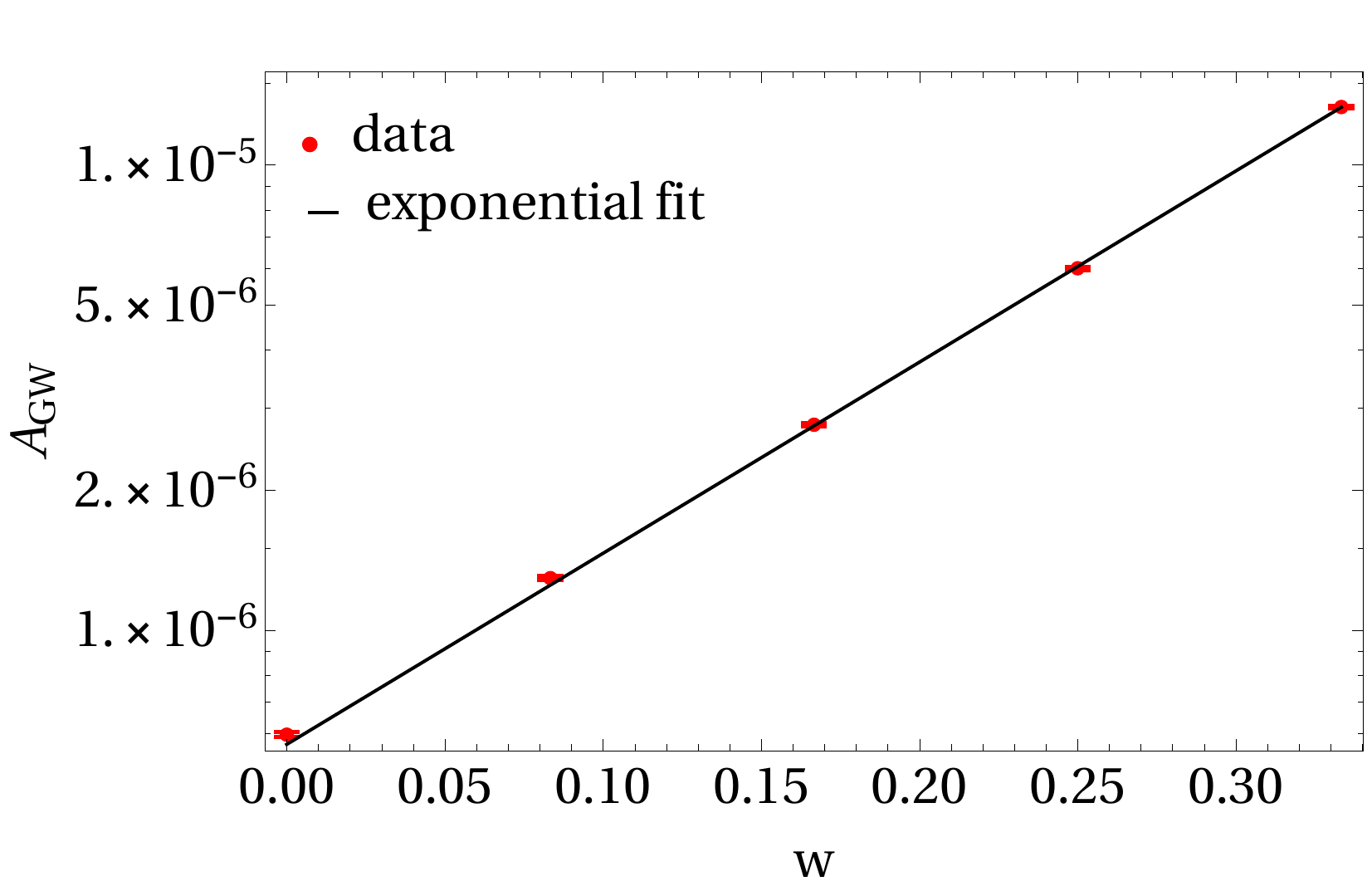}}
\hfill
\subfigure{\includegraphics[width=0.455\textwidth]{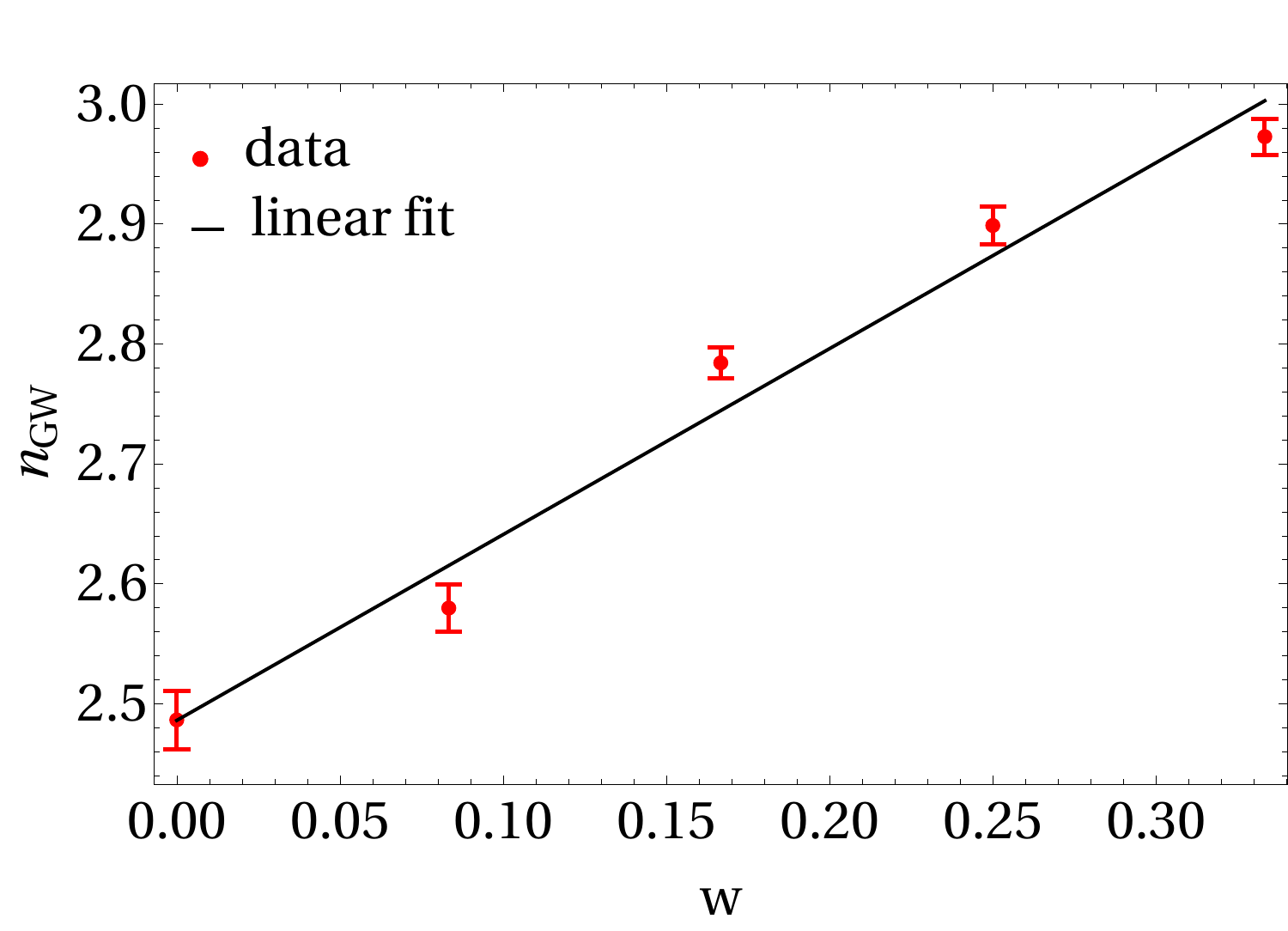}}
\caption{\textit{Left:} The amplitude $A_{\rm GW}$ of eq.~\eqref{eq:Agw} as a function of $w$. The red dots correspond to the data points and their standard error obtained from our fitting analysis (c.f.\ Table~\ref{tab:fitting_analysis}), and the solid black line is an exponential fit to the data.
\textit{Right:} The spectral index $n_{\rm GW}$ as a function of $w$. The data points and their standard error are shown in red and the solid black line corresponds to a linear fit to the data.
} 
\label{fig:fits_different_w}
\end{figure}

To investigate the effect $w$ we numerically computed the GW spectrum, evaluating eq.~\eqref{eq:omega_k_tau_multi} for a single, centered oscillon with background cosmology given by
\be
a(t)=\left(\frac{9}{4}\right)^{\frac{1}{3\,(1+w)}}\left(\frac{2}{3} + H_0\,(1+w)\,t \right)^{\frac{2}{3\,(1+w)}}\,,
\label{eq:aoft_analytic}
\en
such that
\be
a(0) = 1\,,\qquad\textrm{and}\qquad\left.\frac{\dot{a}(t)}{a(t)}\right|_{t=0} = H_0\,.
\en
We computed the spectrum for five different values of $w$ between $w=0$ (matter dominated) and $w=1/3$ (radiation dominated) in steps of $\Delta w=1/12$. The results of our computations are presented in Figures~\ref{fig:spectra_different_w} and~\ref{fig:fits_different_w}. Figure~\ref{fig:spectra_different_w} shows the spectra computed from eq.~\eqref{eq:omega_k_tau_multi} at $a(\tau_f)\equiv a_f = 10$. One can clearly see, that the slope of the peak (on a $\log$-$\log$ scale) increases when increasing the value of $w$. To investigate the $w$-dependence of the peak more closely we assume the following power-law ansatz to fit the computed data:
\be\label{eq:Agw}
\Omega_{\rm GW}(k_{\rm phys}) = A_{\rm GW}\,\left(\frac{k_{\rm phys}}{\omega_{\rm osc}}\right)^{n_{\rm GW}}\,,
\en
where $A_{\rm GW}$ is the amplitude of the peak in the GW spectrum at 
\be
k_{\rm phys}=\omega_{\rm osc}\,,
\en
and 
\be
n_{\rm GW} \equiv \frac{d\log \Omega_{GW}}{d\log k_{\rm phys}}\,,
\en
is the spectral tilt. 

The solid lines in Figure~\ref{fig:spectra_different_w} correspond to the results of our fitting analysis, with details listed in Table~\ref{tab:fitting_analysis}. 
The left part of Figure~\ref{fig:fits_different_w} shows the amplitude $A_{\rm GW}$ as a function of $w$. The red dots correspond to the data points and their standard error obtained from our fitting analysis (see Table~\ref{tab:fitting_analysis}). The solid black line corresponds to an exponential fit to the data points. We find that as a function of $w$ the amplitude of $\Omega_{\rm GW}$ can be described as
\be
A_{\rm GW} = \alpha\, e^{\kappa\,w}\,, 
\en
with
\be
\alpha = 5.68\cdot10^{-7}\pm 1.1\cdot10^{-8}\,,\quad \textrm{and}\quad \kappa= 9.46\pm0.06\,.
\en
The right part of Figure~\ref{fig:fits_different_w} shows the spectral index $n_{\rm GW}$ as a function of $w$. Again, the red dots represent the data and the estimated standard error. To fit the data we assumed a linear ansatz
\be
n_{\rm GW} =  \beta + \gamma \cdot w\,,
\en
finding
\be
\beta = 2.49\pm 0.03\,,\quad \textrm{and}\quad \gamma= 1.55\pm0.14\,.
\en 
In summary we find that $\Omega_{\rm GW}(k)$ is well-described by a power law with a spectral index $n_{\rm GW}$ that scales linearly with $w$.

\begin{table}[ht]
\begin{center}
\begin{tabular}{l|c|c}
$w$             & $A_{\rm GW} \pm \rm{SE}$ & $n_{\rm GW} \pm \rm{SE}$  \\
\hline
\hline
0  & $5.956\cdot10^{-7}\pm7.2\cdot10^{-9}$ & $2.48\pm0.02$   \\
1/12   & $1.292\cdot10^{-6}\pm1.3\cdot10^{-8}$ & $2.58\pm0.02$  \\
1/6     & $2.762\cdot10^{-6}\pm1.8\cdot10^{-8}$ & $2.78\pm0.01$ \\
1/4     & $5.996\cdot10^{-7}\pm4.8\cdot10^{-8}$ & $2.90\pm0.02$   \\
1/3     & $1.332\cdot10^{-5}\pm1.0\cdot10^{-7}$ & $2.97\pm0.02$   \\
\hline 
\end{tabular}
 \caption{Results of the fitting analysis. Estimated values for the amplitude $A_{\rm GW}$ and the spectral index $n_{\rm GW}$ as well as their standard error (SE).}
  \label{tab:fitting_analysis}
\end{center}
\end{table}

\newpage
\subsection{Effects of a variable number of oscillons $N=N(\tau)$}
Another interesting question one may ask is how a time dependent number of oscillons $N(\tau)$ can affect the resulting spectrum of GW. In principle, there are several possibilities that one can consider. In this paper, we restrict ourselves to the following three scenarios:
\begin{enumerate}
\item The number of oscillons is constant: $N=\textrm{cst.}$
\item The number of oscillons increases linearly with conformal time: $N\propto \tau.$
\item The number of oscillons increases with the physical volume of the universe: $N\propto a^3.$
\end{enumerate}

To account for a variable number of oscillons we simply use time dependent Heaviside step functions $\Theta(\tau-\tau_q)$ in the expression for the stress-energy tensor eq.~\eqref{eq:TijTT_multi}. For example, in the case that $N$ scales linearly with $\tau$ we have 
\begin{eqnarray}
T^{\rm TT,\,\rm multi}_{ij}(\textbf{k},\tau)&=&\Lambda_{ij,lm}(\textbf{\^{k}})\,\mathcal{T}_{lm}(\textbf{k},\tau)\sum_{q} \Theta(\tau-\tau_q)\,\Phi^2_q(\tau)\,e^{-i\,\textbf{k}\,\textbf{x}^q}\,
\label{eq:TijTT_multi_timedependentN}
\end{eqnarray}
where
\be
\tau_q = \tau_i + (q-1)\Delta\tau  \quad \textrm{with} \quad \Delta\tau =  \frac{\tau_f-\tau_i}{N(\tau_f)}\,,
\en
where $\tau_i$ and $\tau_f$ are the initial and final (conformal) times and $N(\tau_f)=10$ is the final number of oscillons.

For the scenarios mentioned above we computed the spectrum eq.~\eqref{eq:omega_k_tau_multi} at $a(\tau_f)\equiv a_f =3$, with a final number of oscillons $N(\tau_f)=10$. For the background cosmology we assumed a matter dominated universe ($w=0$). The positions $\textbf{x}^q$ are chosen randomly within the volume $\mathcal{V}$ and the phases $\varphi^q$ of the oscillons are also chosen randomly between $0$ and $2\pi$. In all of the three scenarios we used the same oscillon setup i.e.\ same positions, same phases and same asymmetry $\Delta = 0.1$. 

\begin{figure}[ht]
\centering
\subfigure{\includegraphics[width=0.9\textwidth]{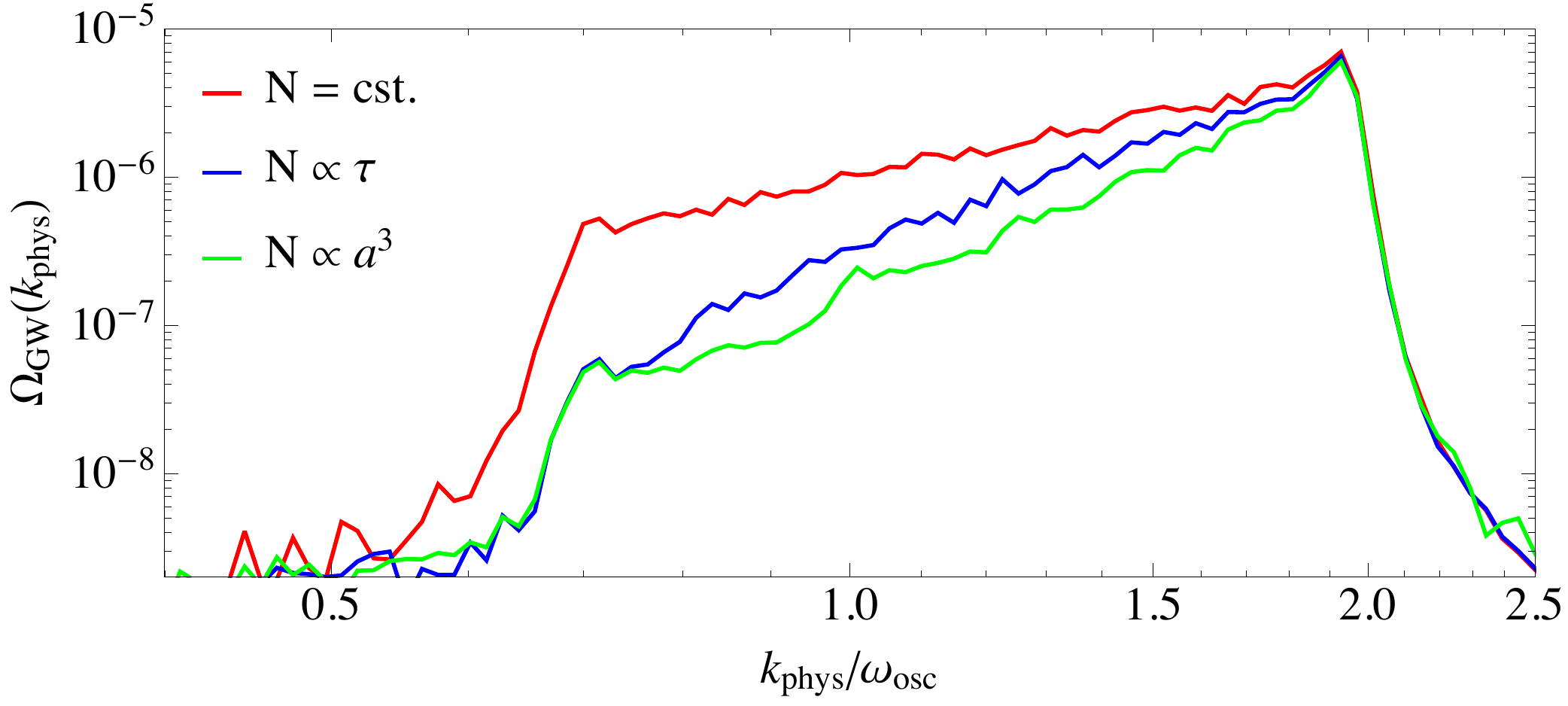}}
\caption{GW spectra computed assuming a matter dominated universe ($w=0$) at $a(\tau_f)\equiv a_f=3$. The three curves represent the following three cases in which: the number of oscillons is assumed to be constant $N=10$ (red), the number of oscillons increases linearly with conformal time $N\propto\tau$ (blue), the number of oscillons increases with the physical volume $N\propto a^3$ (green). In all three cases the final number of oscillons is $N(\tau_f) = 10$.} 
\label{fig:spectra_variable_N}
\end{figure}

Figure~\ref{fig:spectra_variable_N} shows the results of our calculations. The red line corresponds to the spectrum for a constant number of oscillons $N=10$, the blue line shows the resulting spectrum when $N$ increases linearly with conformal time $\tau$, while the green line corresponds to the case where $N\propto a^3$. One can see that the resulting GW spectra clearly differ from each other in all of the three cases. In particular, a crucial difference between the scenario where the number oscillons (per comoving volume) is constant and the other two cases is that the slope of the peak is no more constant on a log-log scale. Furthermore, since the peaks lie perfectly on top of each other for $k_{\rm phys} \gtrsim 2\,\omega_{\rm osc}$ this implies that, ultimately, the amplitude of the GW spectrum does not depend on the evolution of $N$ but rather only on the total number at the time of emission. In other words, the amplitude of the spectrum emitted at $a(\tau)\sim a_f$, i.e.\ when the oscillon distribution is identical in all three cases, is the same in any of the three cases, independently of the previous evolution of $N$. The reason for this is of course that the oscillons continuously produce GWs with the physical wavenumber $\omega_{\rm osc}$, which then get redshifted.

\subsection{Comments on other effects}
We also checked how the GW spectrum $\Omega_{\rm GW}$ scales when changing the number of oscillons per comoving volume ($N=\textrm{cst.}$). We found that in the case of multiple oscillons $N$ the spectrum scales to a good approximation as one would expect, i.e.\ as 
\be
\Omega^{N}_{\rm GW} \sim \Omega^{\textrm{single}}_{\rm GW}\,(N\pm\delta)\,\quad\textrm{with}\quad \delta \lesssim \mathcal{O}(1)\,,
\en
where $\Omega^{\textrm{single}}_{\rm GW}$ corresponds to the spectrum obtained from a single oscillon within the same comoving volume and $\delta$ is a numerical factor depending on the positions $\textbf{x}^q$ and phases $\varphi^q$ of the oscillons. 

To conclude this section we would like to comment on some other effects that we have not considered within our analysis but can in principle have an effect on the resulting GW spectrum. In a more realistic setup, most of the quantities that we assumed to be constant, such as the shape of the oscillons or the amplitude $A$, are of course dynamical and can vary with time. Depending on how these quantities change with time the emission of GWs can be either enhanced or suppressed. Collisions of oscillons, for example, can give rise to large deformations and increased amplitudes resulting in an increased production of GW. On the other hand, at some point the oscillons are expected to start loosing energy. This can lead to a decrease in the amplitude of the oscillons and thus to a reduced production of GW. The final source of GW may be the decay of the oscillons\footnote{For works on the lifetime of oscillons see e.g.\ \cite{Graham:2006xs,Gleiser:2009ys,Amin:2010jq,Saffin:2014yka}.}.

\section{Examples and comparison to lattice simulations}
\label{sec:ex_and_comparison}
In this section we compare results from numerical lattice simulations that were carried out using the program LATTICEEASY \cite{Felder:2000hq} to results that were produced by numerically evaluating eq.~\eqref{eq:omega_k_tau_multi}. We consider two explicit examples: the first example we want to look at is an example of moduli stabilisation in type IIB string theory. To be more precise we consider a realisation of the KKLT scenario \cite{Kachru:2003aw} in which the overall volume modulus is initially displaced from its post-inflationary minimum \cite{Antusch:2017flz}. The second example will be a scenario hilltop inflation which has been extensively discussed in \cite{Antusch:2014qqa,Antusch:2015vna,Antusch:2015ziz,Antusch:2015nla}. 

Although we provide all the necessary information to reproduce the results we will not discuss the models and the associated dynamics in great detail but will rather concentrate on comparing the GW spectra. The phenomenological aspects as well as the field dynamics have been discussed in detail in~\cite{Antusch:2017flz} for the considered KKLT scenario and in~\cite{Antusch:2014qqa,Antusch:2015vna,Antusch:2015ziz,Antusch:2015nla,Antusch:2016con} for the hilltop inflation model. 

In order to evaluate eq.~\eqref{eq:omega_k_tau_multi} in a way that the results can be compared to those from the lattice simulations we proceed as follows:
\begin{enumerate}
\item{\textit{Parameters extracted from lattice simulations:}} The following parameters are extracted from our lattice simulations:
\begin{itemize}

\item The comoving volume $\mathcal{\mathcal{V}}$ simply corresponds to the physical size of the box at the time when stable oscillons start being present in the corresponding lattice simulation: 
\be
\mathcal{V}=\mathcal{V}_{\rm lattice}\cdot a_0^3\,,
\en
where $\mathcal{V}_{\rm lattice}$ is the comoving size of the box and $a_0$ is the scale factor at the time when stable oscillons start to form.

\item The number of oscillons $N$ per comoving volume $\mathcal{\mathcal{V}}$ can be extracted from the three dimensional energy density distribution by counting the large, bubbly energy overdensities. Explicitly, we will count the regions where $\rho/\langle\rho\rangle \ge 16$.
\item The amplitude of the oscillons $A$ is extracted using the field histograms outputted by LATTICEEASY. We assume the amplitude to be somewhat smaller than the maximum amplitude found from the histograms.

\item The width of the oscillons $R$ is extracted by looking at the cross sections of single oscillons. As illustrated in Figure~\ref{fig:osc_profile}, we measure the width of the oscillons at a fraction $1/\sqrt{e}$ of their (current) amplitude. 

\end{itemize}

\begin{figure}[ht]
\centering
\includegraphics[width=0.8\textwidth]{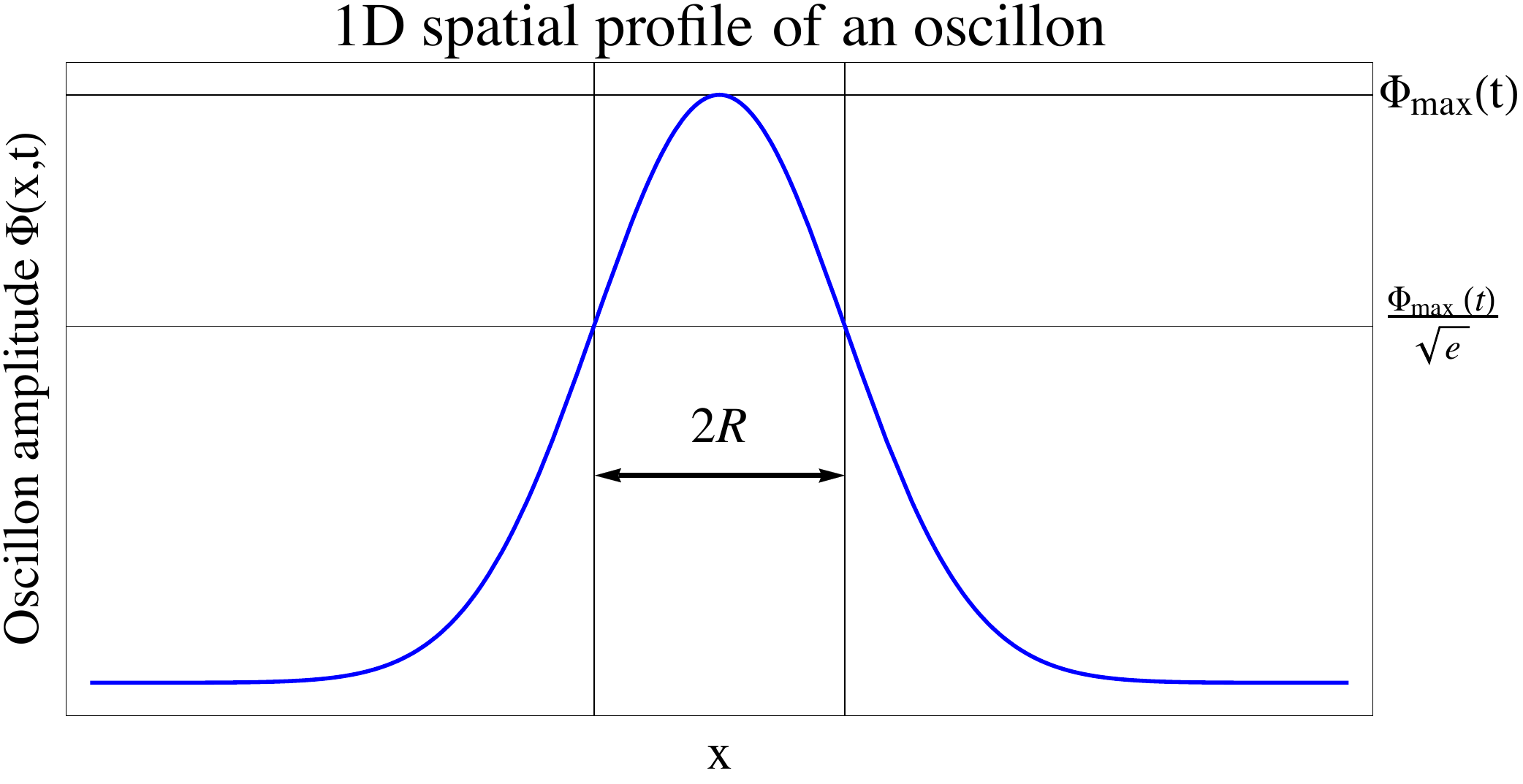}
\caption{Schematic illustration of how the width of an oscillon is extracted.} 
\label{fig:osc_profile}
\end{figure}

\item{\textit{Choosing the background cosmology:}} If not otherwise stated we assume a matter dominated FLRW background, i.e.\ 
\be
a(\tau) \propto \tau^2 \quad \text{with}\quad a(\tau_i)=1 \quad \text{and} \quad a'(\tau_i) = H_0\,
\en
where $H_0$ is the Hubble parameter extracted from the lattice simulations at the time at which stable oscillons are formed.
\item{\textit{Oscillon positions $\mathbf{x}^q$:}} The positions of the oscillons $\mathbf{x}^q$ are randomly chosen within the volume $\mathcal{\mathcal{V}}$ in a way such that the minimum distance between two oscillons $d_{\rm min}$ is not smaller than four times the width of the oscillons:
\be
d_{\rm min} \le 4R\,.
\en
\item{\textit{Oscillon frequency $\omega_{\rm osc}$:}} To model the time dependence of the oscillon amplitude $\Phi_q(t)$ we will again assume a simple cosine:
\be
\Phi_q(t) = A\,\cos(\omega_{\rm osc}t +\varphi^q)\,,
\en
where, as mentioned above, the maximum amplitude $A$ will be extracted from the lattice simulations. The phases $\varphi^{q}$ are randomly chosen between $0$ and $2\pi$:
\be
\varphi^q \sim U([0,2\pi])\,.
\en
Note that both, the KKLT as well as the hilltop potentials are asymmetric around the minimum (see Sections~\ref{subsec:KKLT} and~\ref{subsec:hilltop}). This means that the source will have the same periodicity as the oscillons. For the computation of the GW spectrum via eq.~\eqref{eq:omega_k_tau_multi}, the relevant frequency will be that of the source. Hence, instead of choosing the oscillation frequency of the oscillons $\omega_{\rm osc}$ we will rather choose the oscillation frequency of the source (i.e.\ essentially $2\,\omega_{\rm osc}$) to correspond to the position in $k$-space of the peak in the GW wave spectrum obtained from the lattice simulations. 

\item{\textit{The asymmetry $\Delta$:}} The asymmetry $\Delta$ is chosen within a reasonable range typically
\be
\mathcal{O}(0.1)\lesssim\Delta\lesssim\mathcal{O}(1)\,.
\en
\end{enumerate}

\subsection{Comparison to KKLT}
\label{subsec:KKLT}

\begin{figure}[ht]
\centering
\includegraphics[width=\textwidth]{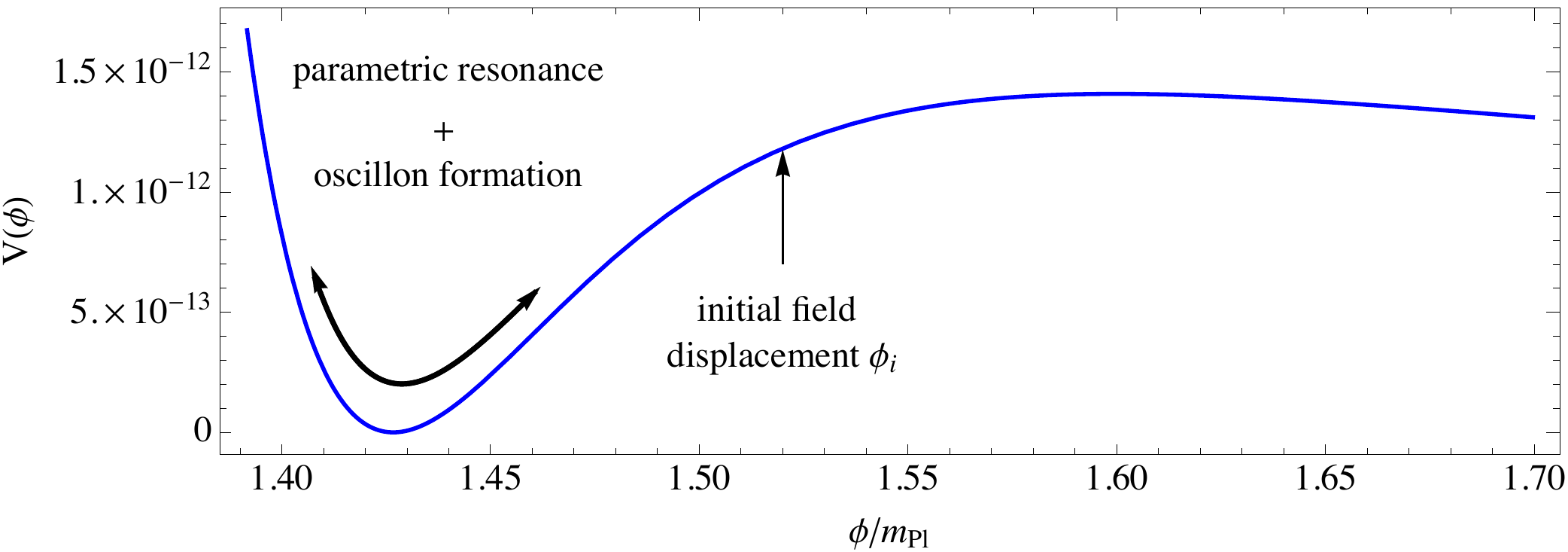}
\caption{The KKLT potential in terms of the canonically normalised field $\phi$. The potential is shown for $W_0=10^{-5}$, $A=10$ and $a=2\pi$.} 
\label{fig:KKLT_potential}
\end{figure}
The first results we want to confront with the semi-analytical method described above originate from numerical lattice simulations of a realisation of the KKLT scenario \cite{Kachru:2003aw}. More precisely we considered the case of a single overall K\"{a}hler modulus which is initially displaced from the minimum of its scalar potential. For the setup considered in this paper the phenomenology and the dynamics have been recently discussed in detail in \cite{Antusch:2017flz}. Here we will only briefly comment on the dynamics and mainly concentrate on comparing the results for the spectrum of GWs. 

Our numerical lattice results were obtained by integrating the scalar field evolution for the canonically normalised K\"{a}hler modulus $\phi$ in its potential
\begin{equation}
V(\phi)/m^4_{\rm Pl} = \frac{2}{3} e^{-\frac{4}{\sqrt{3}}\phi} \left[\frac{a A}{2} e^{-a e^{\frac{2}{\sqrt{3}}\phi}} \left(A \left(a e^{\frac{2}{\sqrt{3}}\phi}+6\right)-6\,W_0\,e^{\frac{1}{2} a e^{\frac{2}{\sqrt{3}}\phi}}\right)+6 D\right]\,.
\label{eq:canonical_potential_KKLT}
\end{equation}
The integration was performed using a modified version of LATTICEEASY \cite{Felder:2000hq} with the parameters $a$, $A$ and $W_0$ fixed to \cite{Antusch:2017flz}
\be
W_0 = 10^{-5}\,,\qquad A =  10\,,\qquad \textrm{and} \qquad  a = 2\pi~.
\label{eq:parameters_KKLT}
\en
The value of $D$ was set to $D=4.68242\times10^{-12}$, in order to uplift the AdS vacuum which is a generic feature of the KKLT model, to a dS minimum. 
The potential~\eqref{eq:canonical_potential_KKLT} with parameters as in~\eqref{eq:parameters_KKLT} is shown in Figure~\ref{fig:KKLT_potential}. 

We consider the case where $\phi$ is initially displaced from the minimum of its potential. When the Hubble parameter $H$ becomes comparable to the mass $m_{\phi}$ of the modulus, the field eventually becomes dynamical and starts oscillating around the minimum of the potential $\phi = \phi_{\rm min}$. In \cite{Antusch:2017flz} we found that these oscillations can lead to the growth of fluctuations of $\phi$ via a parametric resonance and ultimately to the formation of oscillons. 

The lattice results presented below were obtained by simulating the evolution of $\phi$ in its potential eq.~\eqref{eq:canonical_potential_KKLT} on a discrete spacetime lattice in $3+1$ dimensions. The initial conditions for the homogeneous components of the field $\phi$ and its velocity $\dot{\phi}$ were set to
\be
\phi_{\rm i} = 1.53\,m_{\rm Pl}\, \quad \dot{\phi}_{\rm i} =0\,, 
\en
with the initial Hubble parameter purely determined by the potential energy of the field 
\be
H_{\rm i} = \frac{1}{m_{\rm Pl}}\sqrt{\frac{V(\phi_{\rm i})}{3}} = 6.44852\times10^{-7}\,m_{\rm Pl}\,.
\en
\begin{figure}[ht]
\centering
\includegraphics[width=0.9\textwidth]{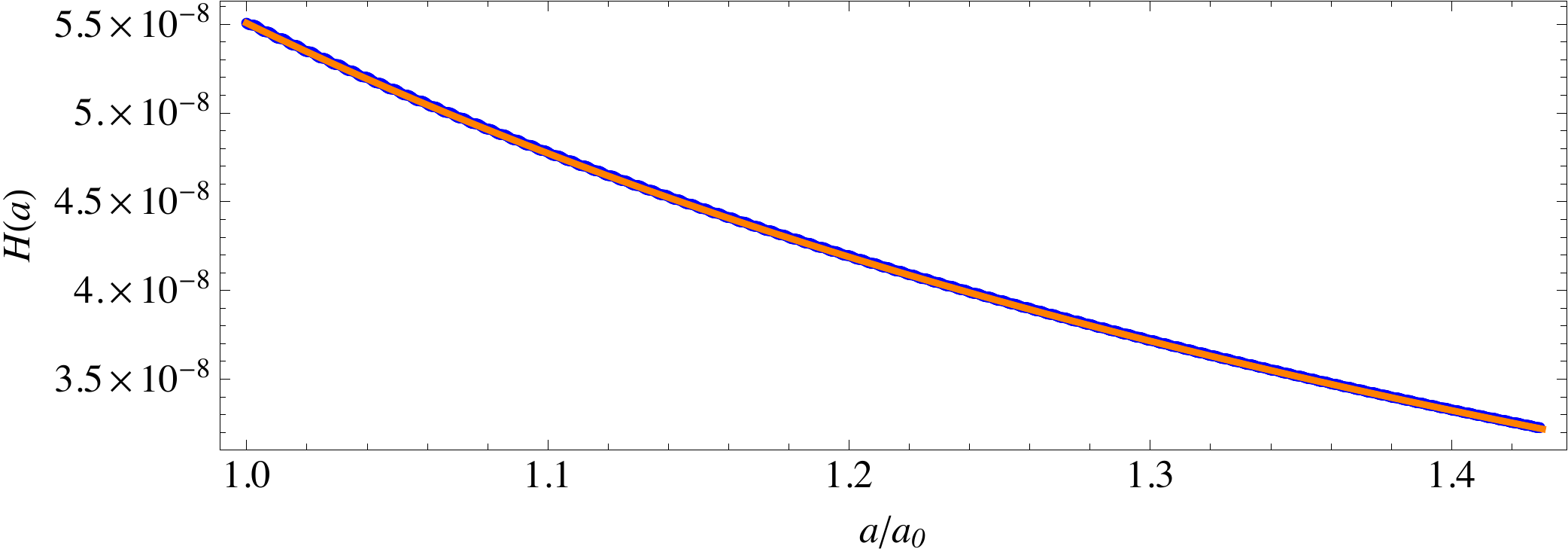}
\caption{Evolution of the Hubble parameter $H$ as a function of the scale factor $a$ normalized to the scale factor at the time of oscillon formation $a=a_0$. The blue dots represent the output of the lattice simulation of the KKLT model, while the orange curve corresponds to assuming a matter dominated universe with initial Hubble parameter equal to the Hubble parameter $H_0$ at the time of oscillon formation $a=a_0$.} 
\label{fig:KKLT_Hofa}
\end{figure}

As usual, the initial field fluctuations are set by vacuum fluctuations\footnote{For more details on how the field fluctuations are initialised on the lattice see \cite{Felder:2000hq}.} \cite{Polarski:1995jg,Khlebnikov:1996mc}.
The simulation was performed in a box of comoving size $\mathcal{V}_{\rm lattice}=L^3\simeq (0.7/H_0)^3$ with 512 lattice points per spatial dimension.

We ran our lattice simulation up to a scale factor $a_{\rm end}\simeq8.3$ and found that oscillons start to form at $a_0\simeq5.8$. Thus, the relative expansion between the formation of oscillons and the end of our simulations corresponds effectively to a factor $a_f \equiv a_{\rm end}/a_0\simeq 1.38$. In Figure~\ref{fig:KKLT_Hofa} we show the Hubble parameter as a function of the scale factor $a/a_0$ for $a_i=1\le a/a_0\le a_{f}$. The blue dots correspond to the output of our lattice simulation. The orange curve corresponds to assuming a matter dominated universe i.e.\ essentially to eq.~\eqref{eq:aoft_analytic} with $w=0$. The value of $H_0$ in eq.~\eqref{eq:aoft_analytic} was fixed by taking the value of the Hubble parameter from the lattice simulation at $a=a_0=5.8$. One can see that the cosmological evolution on the lattice is in excellent agreement with a matter dominated universe. 

As mentioned above we extracted some of the parameters (such as the amplitude or the width of the oscillons) from the lattice simulation to compute the spectrum of GWs via eq.~\eqref{eq:omega_k_tau_multi}. How this was done in practice is explained in Appendix~\ref{sec:parameters_KKLT}. The values of the parameters that we used for the numerical evaluation of~\eqref{eq:omega_k_tau_multi} are summarized in Table~\ref{tab:setup_KKLT}. Interestingly, we found that according to the lattice simulation the number of oscillons $N$ increases with time. In our semi-analytical calculation of $\Omega_{\rm GW}$ we have taken this fact into account by assuming that the number of oscillons per \textit{physical} volume remains constant. Explicitly we have used (see Appendix~\ref{sec:parameters_KKLT}):
\be
N(a) = N(a_0)\cdot\left(\frac{a}{a_0}\right)^3\,\quad \textrm{with}\quad N(a_0) = 4\,.
\en
With $a_0/a_{\rm end}\simeq1.43$ we have a final number of oscillons of $N(a_{\rm end}) \simeq 12$. 
For the background cosmology we assumed matter dominated universe with (cf.\ Figure~\ref{fig:KKLT_Hofa})
\be
a(t) =  \left(\frac{3}{2}\right)^{2/3}\,\left(\frac{2}{3} + H_0 t\right)^{2/3}\,\quad\textrm{and}\quad H_0= 5.2227\times10^{-8}\,m_{\rm Pl}\,.
\en
The asymmetry $\Delta$ was considered as a free parameter. 

\begin{figure}[ht]
\centering
\subfigure{\includegraphics[width=0.9\textwidth]{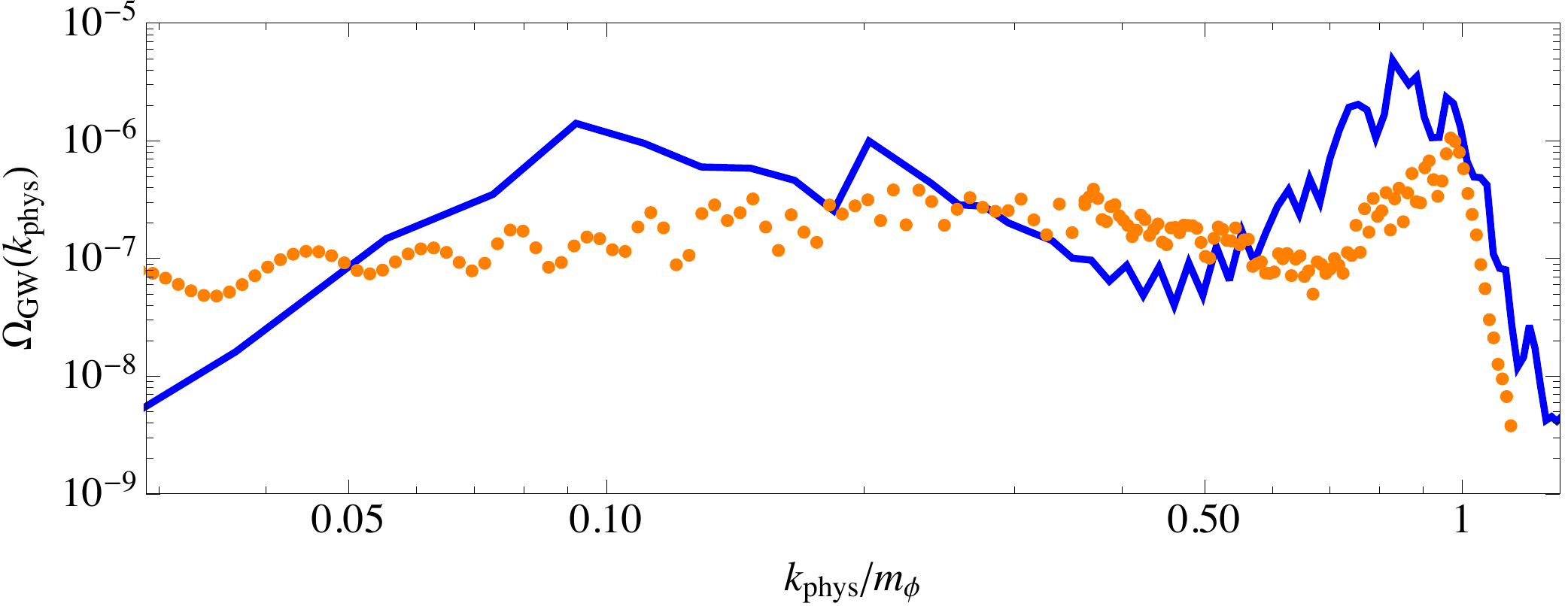}}
\caption{Semi-analytical result of the GW spectrum $\Omega_{\rm GW}$ (orange dots) vs.\ the final spectrum obtained from the lattice simulation of the KKLT model (solid blue curve). The spectrum is shown as a function of the physical wavenumber $k_{\rm phys}/m_{\phi} = a_f^{-1} k/m_{\phi}$, where $m_\phi$ is the mass of the modulus at the minimum of the potential.} 
\label{fig:Omega_KKLT_comparison}
\end{figure}

In Figure~\ref{fig:Omega_KKLT_comparison} we present the GW spectrum at the end of the lattice simulation (solid blue line) compared to the semi-analytical result obtained from numerically evaluating eq.~\eqref{eq:omega_k_tau_multi} (orange dots). The spectra are shown as a function of the physical wavenumber $k_{\rm phys}/m_{\phi} = a^{-1} k/m_{\rm \phi}$. 
We find that our semi-analytical results are in good agreement with the results from the lattice simulation when assuming
\be
\Delta = 0.5\,.
\en
The differences between the semi-analytical approach and the lattice simulation can be caused by various effects.
For example, we assumed that all oscillons have the same size which is not necessarily the case (see Appendix~\ref{sec:parameters_KKLT}). Differences in the slope of the peak can arise if the number of oscillons $N$ scales differently than $N\propto a^3$, as assumed in our semi-analytical computation. Moreover, we note that interference effects from the overlap of the profiles of very close oscillons are not taken into account by our semi-analytical approach (see Section~\ref{sec:multiple_osc}). Particularly in this scenario, where oscillons are sometimes observed to lie very close to each other, neglecting this interference contribution may result in an underestimation of the emitted GWs.

\begin{table}
\begin{center}
\begin{tabularx}{12cm}{L|L|L|L}
oscillon amplitude $A$ &  oscillon width $R$ & number of oscillons $N$ & comoving volume $\mathcal{V}$ \\
\hline
\hline
$0.0085\,m_{\rm Pl}$ & $99234.9\,m^{-1}_{\rm Pl}$  & $N(a) \propto \,a^3$,  $N(a_0) = 4$, $N(a_{\rm end}) = 12$  & $2.574\times10^{20}\,m^{-3}_{\rm Pl}$ \\

\end{tabularx}
 \caption{Parameters extracted from the lattice simulation of the KKLT model that were used for the semi-analytical computation of the GW spectrum.}
  \label{tab:setup_KKLT}
\end{center}
\end{table}

\subsection{Comparison to hilltop inflation}
\label{subsec:hilltop}
\begin{figure}[ht]
\centering
\includegraphics[width=0.9\textwidth]{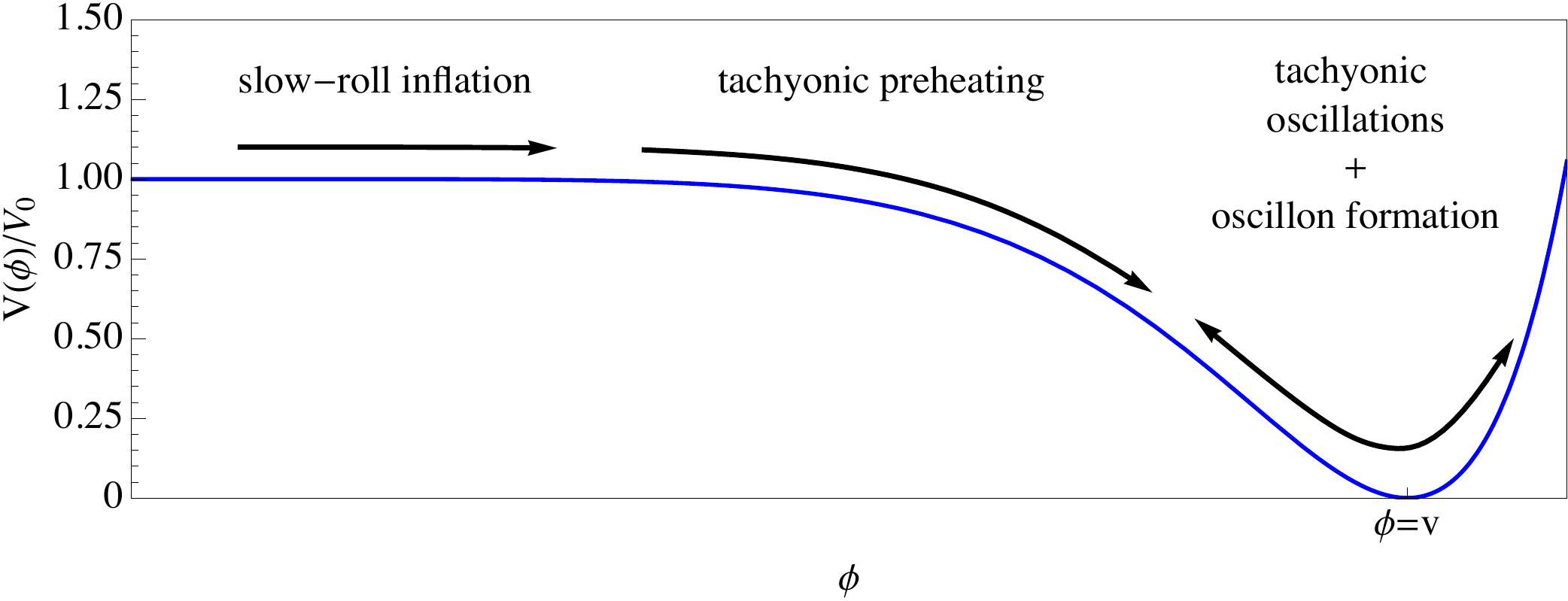}
\caption{Schematic plot of the hilltop potential in unit of $V_0$. The potential is shown for $\phi \ge 0$. The different phases of the field evolution are schematically depicted by the black arrows.} 
\label{fig:hilltop_potential}
\end{figure}
For our second and last comparison we consider a model of (small field) hilltop inflation which is characterised by the following potential
\be
V(\phi) = V_0\left(1 - \frac{\phi^6}{v^6} \right)^2\,,
\label{eq:hilltop_potential}
\en
where $V_0$ is a constant vacuum energy. The potential exhibits two minima at $\phi=\pm v$, which are separated by a flat plateau. In Figure~\ref{fig:hilltop_potential} we show the potential eq.~\eqref{eq:hilltop_potential} in units of $V_0$ for $\phi\ge0$. In what follows, the vacuum energy $V_0$ and the vacuum expectation value $v$ of $\phi$ will be set to
\be
V_0 = 1.05\times10^{-19}\,m^4_{\rm Pl},\quad\textrm{and}\quad v= 10^{-2}\,m_{\rm Pl}\,.
\en
While $v$ is chosen arbitrarily with $v\ll m_{\rm Pl}$, the value of $V_0$ is simply chosen in order to be consistent with the most recent Planck bounds on the scalar amplitude $\As \simeq 2.2\times10^{-9}$ (see Refs~\cite{Antusch:2014qqa,Antusch:2015vna,Antusch:2015ziz,Antusch:2015nla} for more details).

As indicated in Figure~\ref{fig:hilltop_potential}, inflation occurs while the inflaton $\phi$ rolls slowly along the flat region of its potential ($\phi \sim 0$) towards one of its minima $\phi=\pm v$. While the inflaton comes closer to its minimum it thereby gains kinetic energy and eventually exits the slow-roll regime. Inflation is subsequently followed by a period of preheating which can be subdivided into two qualitatively different phases: A perturbative phase of tachyonic preheating which leads to a growth of all infrared modes that acquire an effective tachyonic mass squared ($V''(\phi) + k^2/a^2 < 0$), and a subsequent phase of tachyonic oscillations which leads to the amplification of fluctuations around a certain peak scale, somewhat below the mass scale. The growth during this phase is efficient enough to give rise to non-perturbative effects such as the formation of oscillons. 

In this scenario, the mechanism that causes the growth of $\phi$ fluctuations and ultimately leads to the formation of oscillons is qualitatively different than in the KKLT scenario. Furthermore, not only the formation but also the subsequent evolution differs significantly from the one in the KKLT model. We will see in the following, that the different dynamics are also reflected in the emitted spectrum of GWs.

Using LATTICEEASY \cite{Felder:2000hq} we simulated the scalar field evolution of $\phi$ in its potential~\eqref{eq:hilltop_potential}. The simulation was performed in three spatial dimensions with $512^3$ lattice points. The initial conditions for the the field $\phi$ and its time derivative $\dot{\phi}$ were set to
\be
\phi_{\rm i} = 8\times 10^{-4}\,m_{\rm Pl}\, \quad \dot{\phi}_{\rm i} =2.49\times10^{-13}\,m^2_{\rm Pl}\,, \quad\textrm{and}\quad H_{\rm i} = 1.87\times10^{-10}\,m_{\rm Pl}\,,
\en
and the fluctuations were agaian initialised as vacuum fluctuations \cite{Polarski:1995jg,Khlebnikov:1996mc}. 
In units of the initial Hubble parameter, the comoving volume covered by the simulation is $\mathcal{V}_{\rm lattice}=L^3\simeq (\pi/200\, H^{-1}_{\rm i})^3$.

\begin{figure}[ht]
\centering
\includegraphics[width=0.9\textwidth]{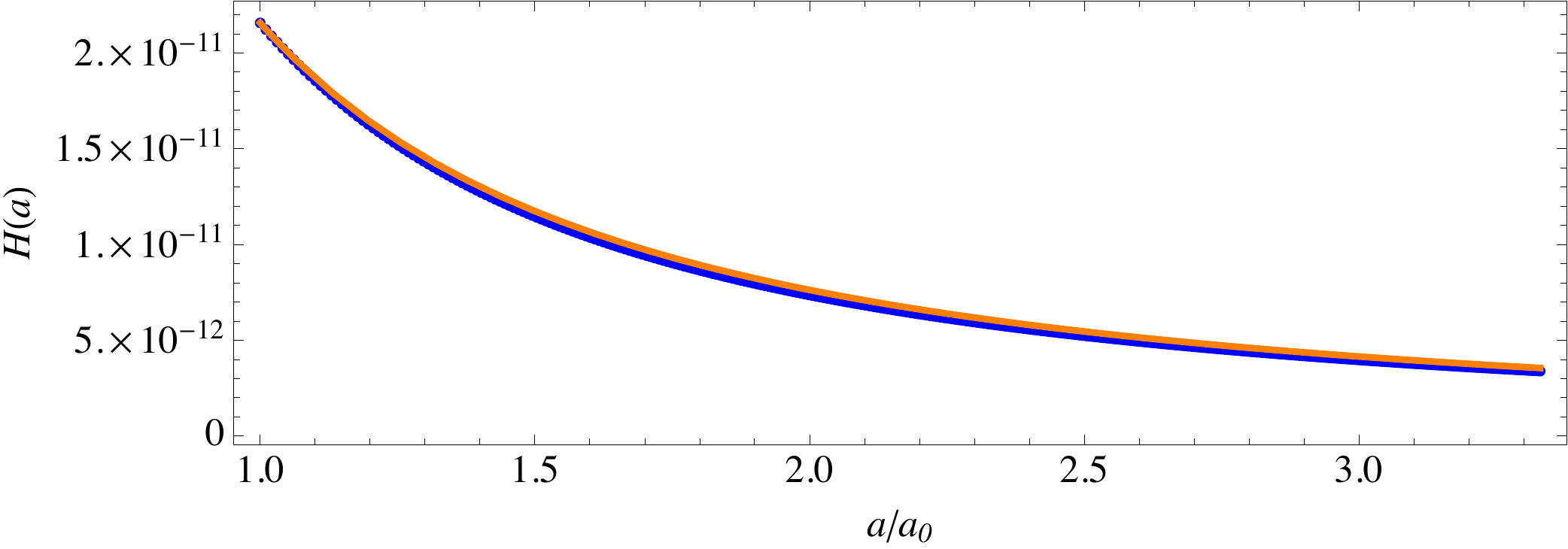}
\caption{Evolution of the Hubble parameter in terms of the scale factor $a$. The evolution is shown in terms of the relative expansion between the time of oscillon formation at $a=a_0$ and the end of the lattice simulation simulation $a=a_{\rm end}$. The blue dots correspond to the output of the lattice simulation of the hilltop inflation model and the solid orange curve corresponds to assuming a matter dominated universe.} 
\label{fig:hilltop_Hofa}
\end{figure}

\begin{table}
\begin{center}
\begin{tabularx}{12cm}{L|L|L|L}
oscillon amplitude $A$ &  oscillon width $R$ & number of oscillons $N$ & comoving volume $\mathcal{V}$ \\
\hline
\hline
$0.4\,v$ & $1.24\times10^7\,m^{-1}_{\rm Pl}\,$  & $N=4=\textrm{cst.}$   & $7.63\times10^{25}\,m^{-3}_{\rm Pl}$ \\

\end{tabularx}
 \caption{The values of the oscillon parameters used for the semi-analytical computation of the GW spectrum $\Omega_{\rm GW}(k_{\rm phys})$ for the hilltop inflation setup. They have been extracted from the numerical lattice simulation.}
  \label{tab:setup_hilltop}
\end{center}
\end{table}
In our simulation of the hilltop model~\eqref{eq:hilltop_potential}, the formation of stable oscillons starts at $a_0\simeq 5$. We ran the simulation up to a scale factor $a_{\rm end} \simeq 16.8$. The relative expansion factor between the time of formation $a_0$ and the end of the simulation is therefore $a_f \equiv a_{\rm end}/a_0\simeq3.3$. Figure~\ref{fig:hilltop_Hofa} shows the evolution of the Hubble parameter between $a=a_0$ and $a=a_{\rm end}$. The Hubble parameter is plotted as a function of the scale factor $a/a_0$. The blue dots correspond to the output of the lattice simulation, while the solid orange curve denotes the Hubble parameter assuming a matter dominated universe
with 
\be
a(t) =  \left(\frac{3}{2}\right)^{2/3}\,\left(\frac{2}{3} + H_0 t\right)^{2/3}\,\quad\textrm{and}\quad H_0= 2.15683\times10^{-11}\,m_{\rm Pl}\,.
\en
The value of $H_0$ corresponds to the value of the Hubble parameter in our lattice simulation at $a=a_0$. In contrast to the KKLT scenario, we observe that the number of oscillons $N$ remains constant over time. From the output of LATTICEEASY we find that in  case the number of oscillons is
\be
N=4=\textrm{cst.}\,. 
\en
The other parameters that we deduced from the lattice simulation are summarized in Table~\ref{tab:setup_hilltop} (see Appendix~\ref{sec:parameters_hilltop} for more details).

\begin{figure}[ht]
\centering
\subfigure{\includegraphics[width=0.9\textwidth]{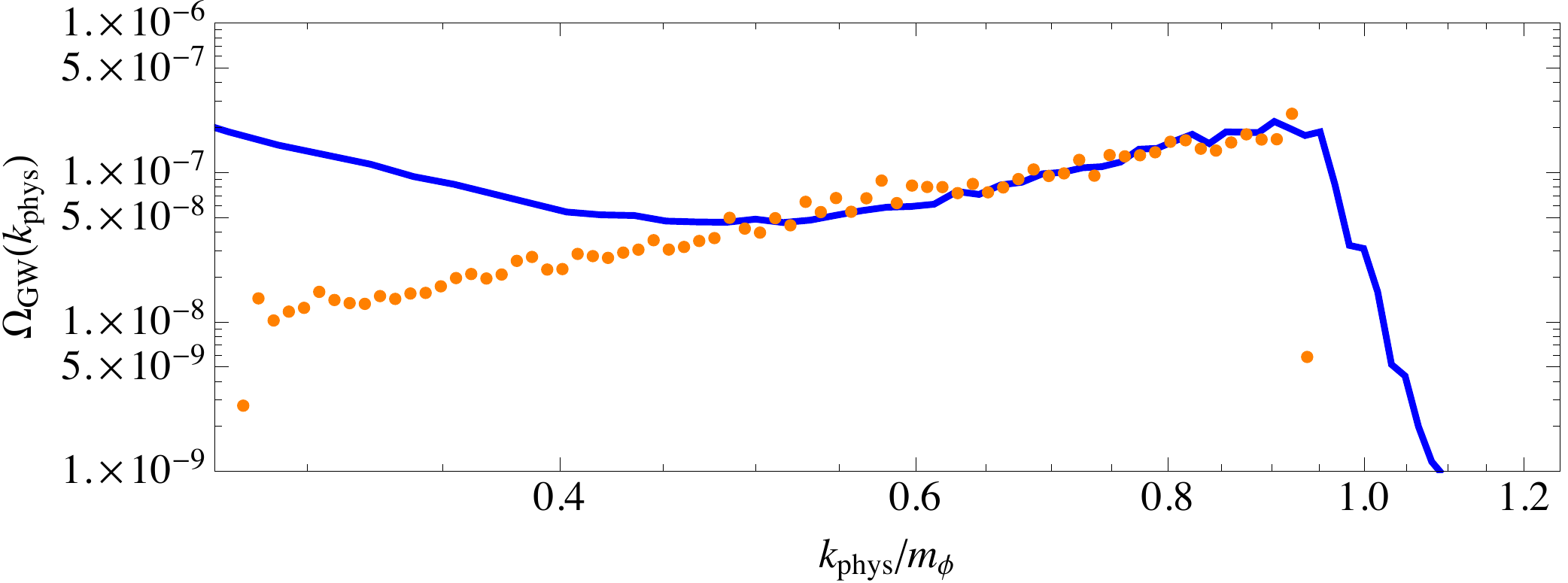}}
\caption{The spectrum of GW as funktion of the physical momentum $k_{\rm phys} = a_f^{-1} k$ over the mass $m_\phi$ of the inflaton at the minimum of the potential. The solid blue line corresponds to the final output of our lattice simulation. The orange dots represent the result of the semi-analytical calculation of $\Omega_{\rm GW}(k)$.}
\label{fig:Omega_hilltop_comparison}
\end{figure}
The result from our semi-analytical computation of the $\Omega_{\rm GW}$ for the hilltop setup is presented in Figure~\ref{fig:Omega_hilltop_comparison} (orange dots) together with the spectrum at the end of the lattice simulation (solid blue line). The spectra are shown as a function of the physical wave number $k_{\rm phys}/m_{\phi} = a_f^{-1} k/m_{\rm \phi}$. We find that the results are in excellent agreement with those from the lattice simulation when fixing the asymmetry $\Delta$ to
\be
\Delta = 0.15\,.
\en
We note that compared to our earlier lattice simulation result for the GW spectrum from an oscillon system in \cite{Antusch:2016con} with $128^3$ lattice points, the peak in the simulation presented here is less pronounced. Comparing the oscillon systems shows that in the new simulation less oscillons were produced which are less asymmetric and also oscillate with somewhat lower amplitudes. These effects can in principle be caused by the initial seed for the simulation, which can have an effect here since in order to maximise resolution of the oscillons we have to choose a rather small box and therefore seed effects might not have averaged out. On the other hand, it may also be caused by a lack of resolution in the $128^3$ simulation (or by a combination of both). In any case, given the different oscillon parameters, the difference between the two simulations can be understood from the analytic expressions. Assuming that the oscillon parameters stay constant, the GW peak would reach the sensitivity of aLIGO run 5 (as shown in Figure 3 of \cite{Antusch:2016con}) when the universe expanded by an additional factor of $\lesssim 10$.

Compared to the KKLT model, it seems that our semi-analytical approach is better suited for a setup similar to that of hilltop inflation. In the hilltop inflation model, the oscillon dynamics is less complex, in the sense that the number of oscillons per \textit{comoving} volume remains constant over time, rendering also interference effects less likely. In contrast to the KKLT model, however, our approach fails to reproduce the spectrum in the IR (i.e.\ $k_{\rm phys}\ll m_{\phi}$) (c.f.\ Figure~\ref{fig:Omega_hilltop_comparison}). This, however, is not related to our approach, but rather to the fact that the IR spectrum in hilltop inflation originates from an early phase of preheating, which takes place before and during  the oscillons are formed.

\section{Summary and Conclusions}
\label{sec:summary_and_conclusions}
In this paper we studied the GW production from asymmetric oscillons using a semi-analytical approach. To calculate the spectrum of GW produced by asymmetric oscillons in an expanding universe we adopted the formalism discussed in \cite{Dufaux:2007pt}, which we briefly reviewed in Section~\ref{sec:preh_dynamics}. Under simplifying assumptions we derived an analytic formula for the transverse-traceless part of the energy-momentum tensor (i.e.\ the source of the GWs) for a system of $N$ randomly distributed, asymmetric oscillons that oscillate with different, random phases. The latter was used to numerically compute the spectrum of GWs. 

In Section~\ref{seq:effects_of_parameters} we studied how the different parameters that characterise an oscillon system can affect the emission of GW. Under the assumption that all oscillons are identical, i.e.\ that they share the same periodicity, same maximum amplitude, same spatial profile and have the same amount of asymmetry, we found that changes in the amplitude as well as in the asymmetry lead simply to an overall rescaling of the GW spectrum. On the other hand, we found that different background cosmologies lead to different growth rates of the characteristic oscillon peak in the GW spectrum (i.e.\ essentially to different constant slopes of the peak on a $\log$-$\log$ scale). The fact that the background cosmology is clearly imprinted in the GW spectrum implies that the observation of such a signal, at some point in the future, could provide valuable information about the equation of state of the universe during its early stages. We also studied the effect of a time-dependent number of oscillons (per comoving volume) as observed in certain models. We found that this leads to a non-constant slope of the peak on a $\log$-$\log$ scale. 

To test the reliability of the semi-analytical method we compared the results to those from numerical lattice simulations. We explicitly tested two different example models which show qualitatively different dynamics. As a first example we considered a realisation of the KKLT model \cite{Kachru:2003aw} (as in \cite{Antusch:2017flz}) and found good agreement between the semi-analytical and the purely numerical results. The second example we considered was a model of hilltop inflation (as in \cite{Antusch:2016con}). Particularly in this case where the number of oscillons (per comoving volume) remains constant, we found an excellent agreement between the semi-analytical approach and the lattice simulation. 

In summary, we found that the semi-analytic techniques are a useful and computationally inexpensive tool for estimating the GW production from oscillons and for studying  the dependence of the GW spectrum on the characteristics of oscillon systems. Our examples highlight that different models can give rise to different such characteristics which manifest themselves in the stochastic background of GWs. Our analysis provides a first step towards exploring what one can learn from the GW spectrum produced by oscillon systems.

\section*{Acknowledgements}
This work has been supported by the Swiss National Science Foundation.

\appendix
\section{Parameter extraction from lattice simulations}
\label{sec:parameters_from_lattice}
In Section~\ref{sec:ex_and_comparison} we compared the GW spectra obtained from numerical lattice simulations to those obtained with our semi-analytical approach. We did this for two explicit models in which the dynamics are qualitatively different. In order to compare our semi-analytical results to the results of the lattice simulations we used some results of the lattice simulations to design an artificial but qualitatively similar oscillon setup to the setup in the corresponding lattice simulations. In what follows, we briefly describe how we extracted some of the parameters characterising an oscillon setup within our semi-analytical approach.

\subsection{Parameter extraction for the KKLT scenario}
\label{sec:parameters_KKLT}
To extract the number of oscillons between the time of formation $a=a_0$ and the end of the simulation $a=a_{\rm end}$ we counted the bubbly regions with 16 times the average energy density. The reason why we considered the three dimensional energy density distributions rather than the field distributions is simply to avoid that some oscillons may be overlooked. Since the oscillons oscillate quickly in time it may be possible that the amplitude of some oscillons is comparable to the amplitude of the background at the time when the output is generated. In such a case the oscillons may be missed by looking at the field distribution, but not by looking at the energy density distribution since the energy within an oscillon is practically conserved on short time scales. 

\begin{figure}[h]
\centering
\subfigure{\includegraphics[width=0.45\textwidth]{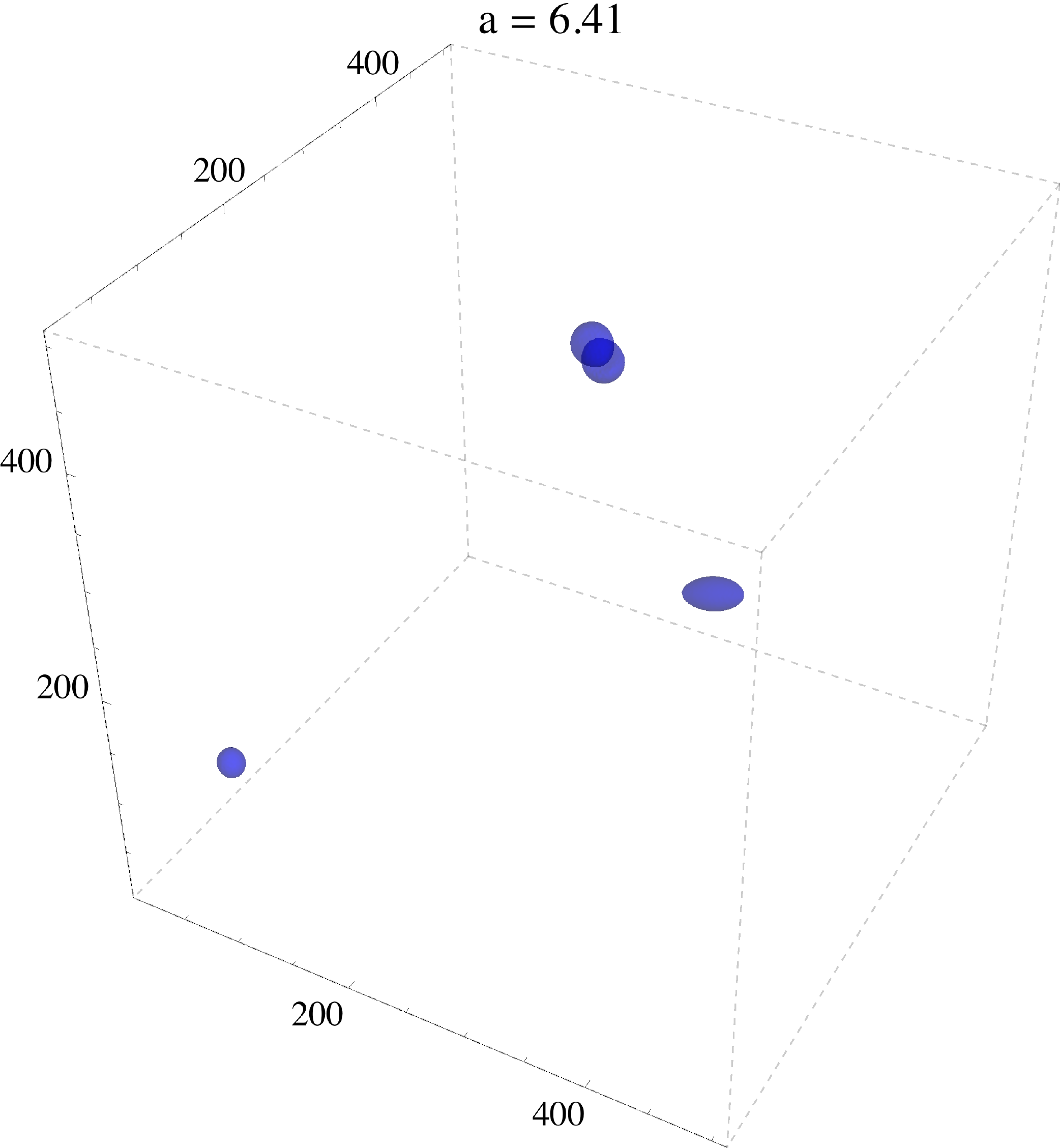}}
\hfill
\subfigure{\includegraphics[width=0.45\textwidth]{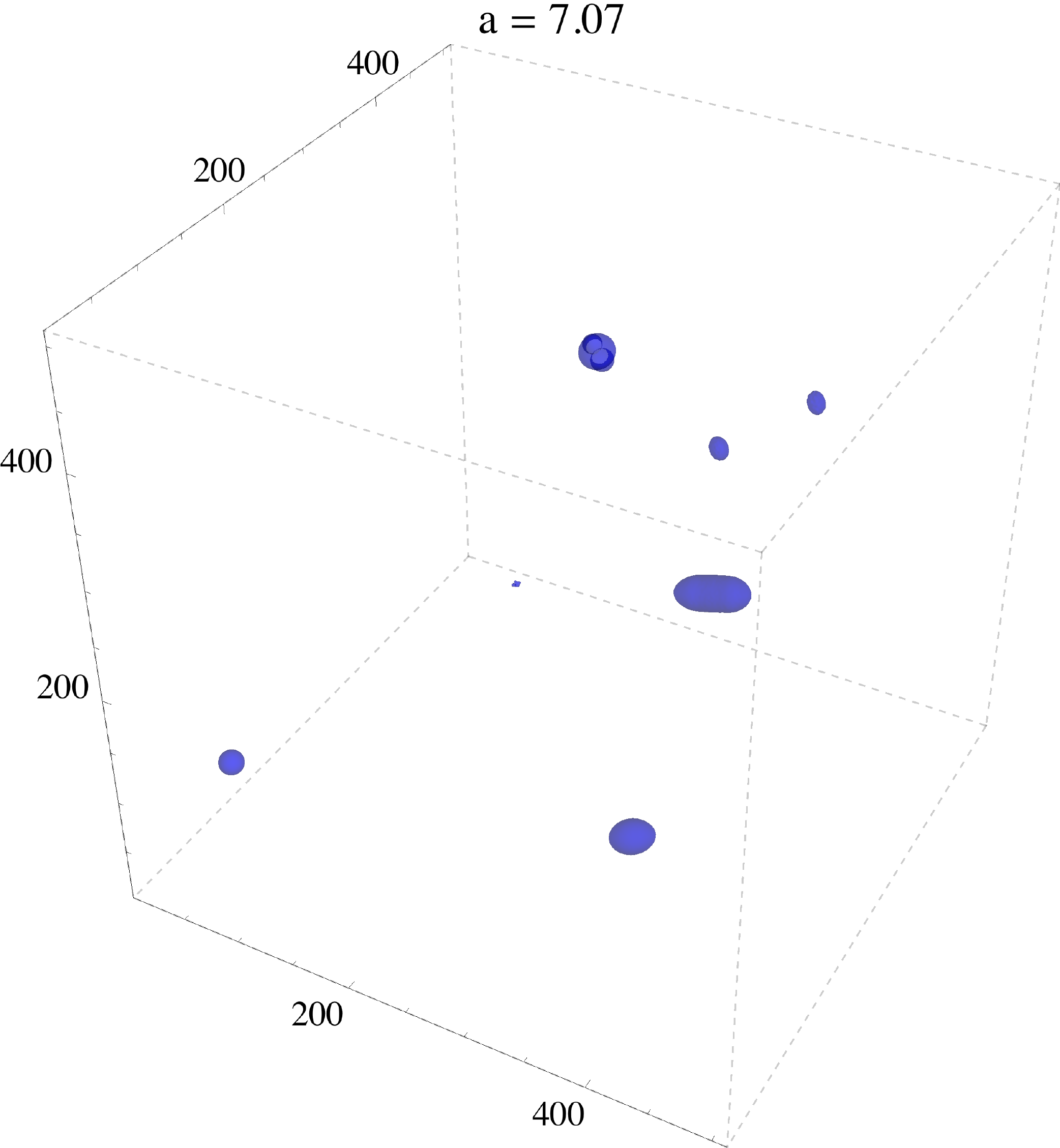}}
\hfill
\subfigure{\includegraphics[width=0.45\textwidth]{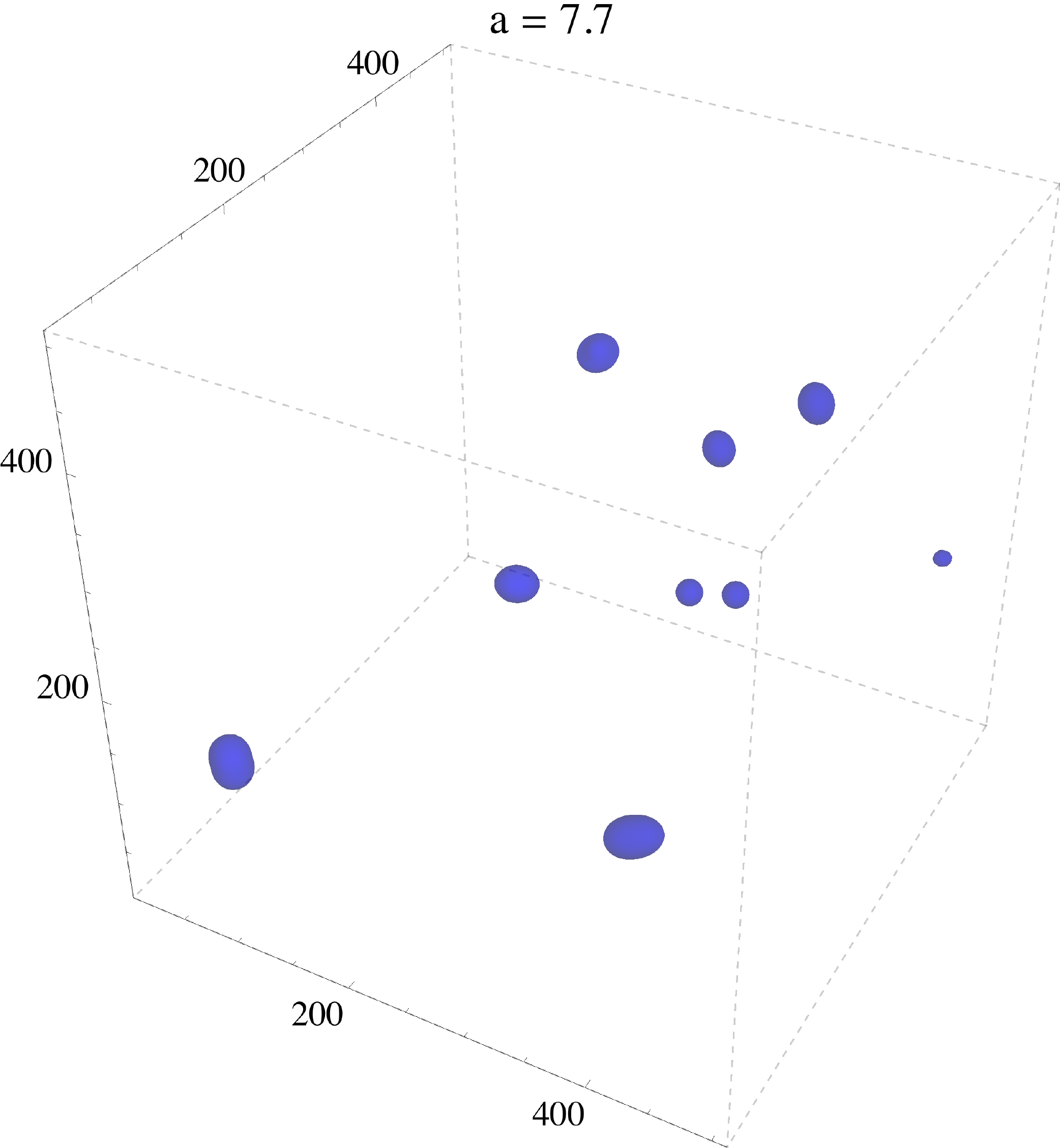}}
\hfill
\subfigure{\includegraphics[width=0.45\textwidth]{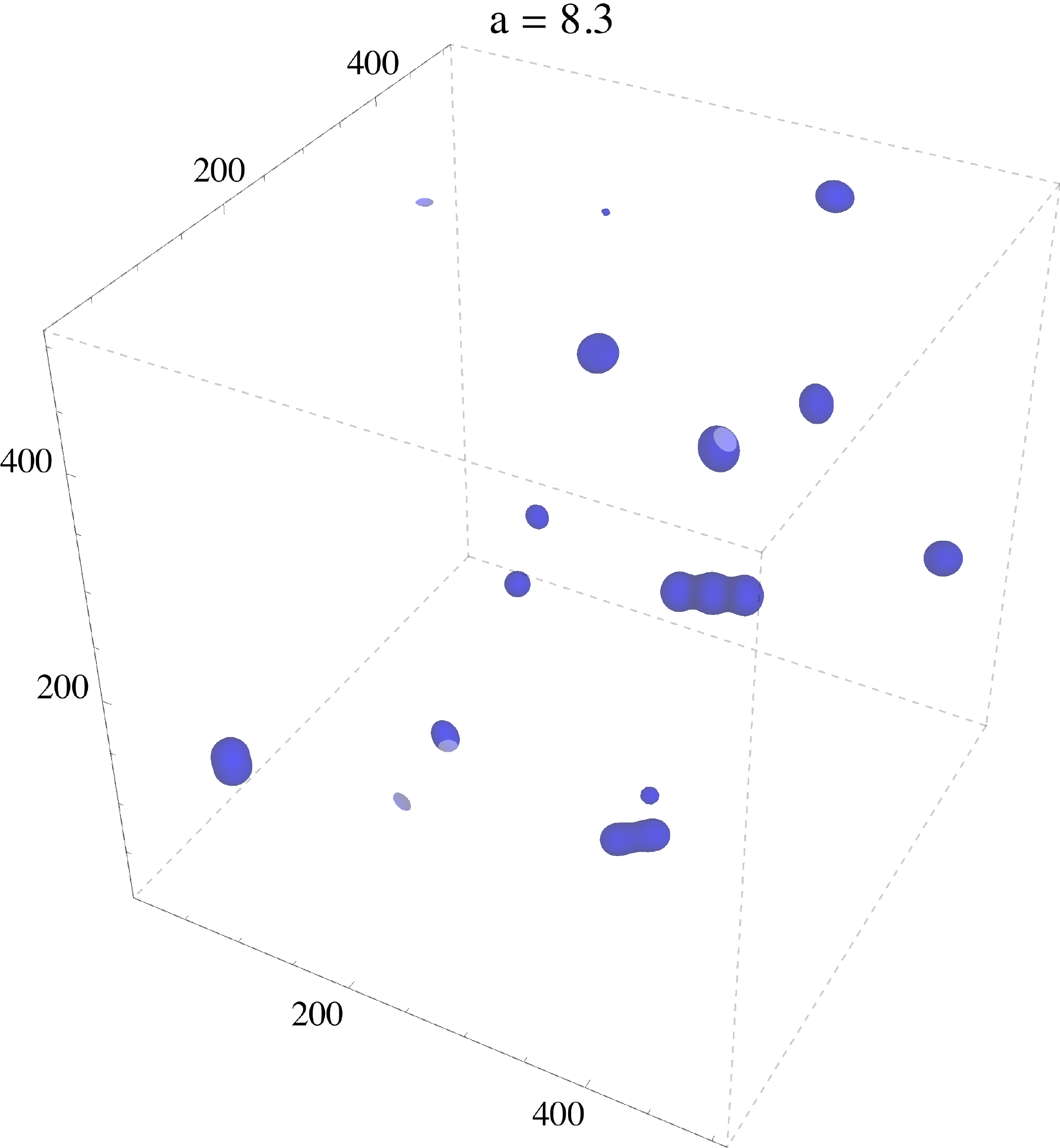}}
\hfill
\caption{Energy density distribution at different moments in time represented by the scale factor $a$. The blue surfaces correspond to regions with $\rho/\langle\rho\rangle \ge 16$. One can see that the more the universe expands the more highly energetic regions, i.e.\ oscillons appear. The figures originate from the lattice simulation of the KKLT model with $512^3$ lattice points.} 
\label{fig:eden_slices_KKLT}
\end{figure}

Figure~\ref{fig:eden_slices_KKLT} shows the energy density distribution in our lattice simulation of the KKLT model. The distribution is shown at different moments in time represented by the scale factor of the lattice simulation. The blue surfaces correspond to regions with 16 times the average energy density $\langle\rho\rangle$. These regions correspond to regions where the field oscillates with a large amplitude compared to its surrounding, i.e.\ to oscillons. One can see that the number of highly energetic regions increases with time. 

\begin{figure}[ht]
\centering
\subfigure{\includegraphics[width=0.45\textwidth]{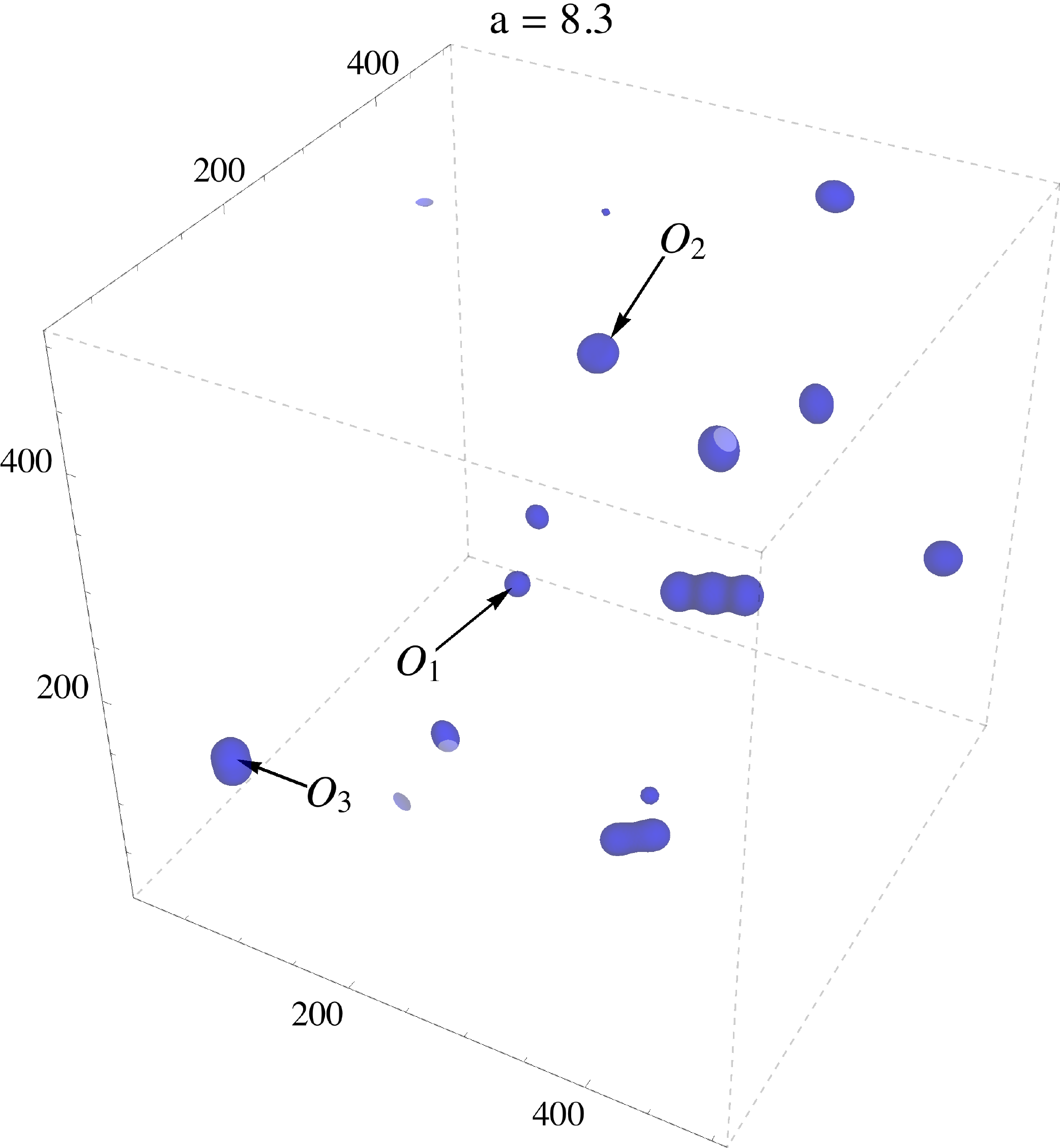}}
\hfill
\subfigure{\includegraphics[width=0.45\textwidth]{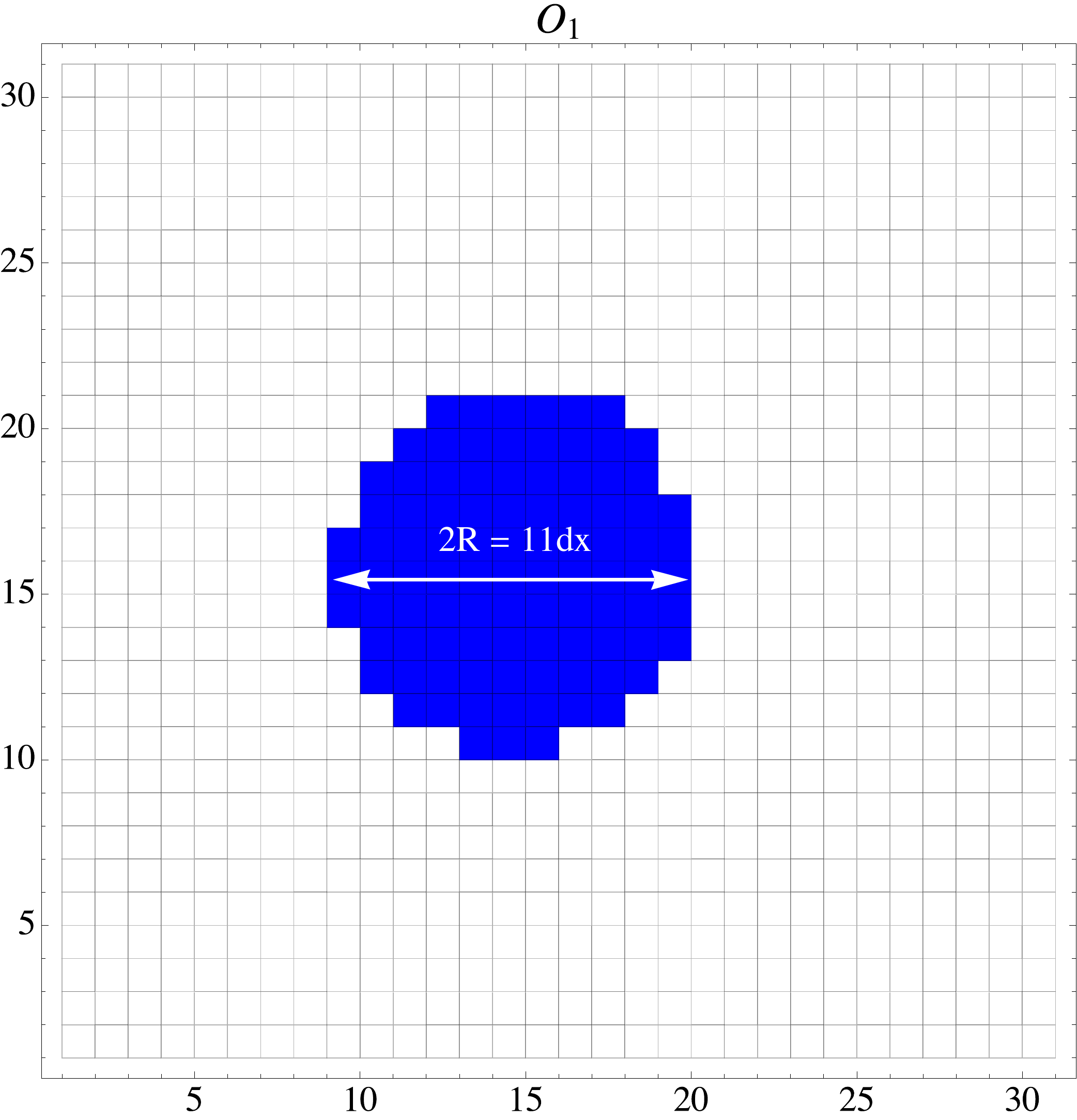}}
\hfill
\subfigure{\includegraphics[width=0.45\textwidth]{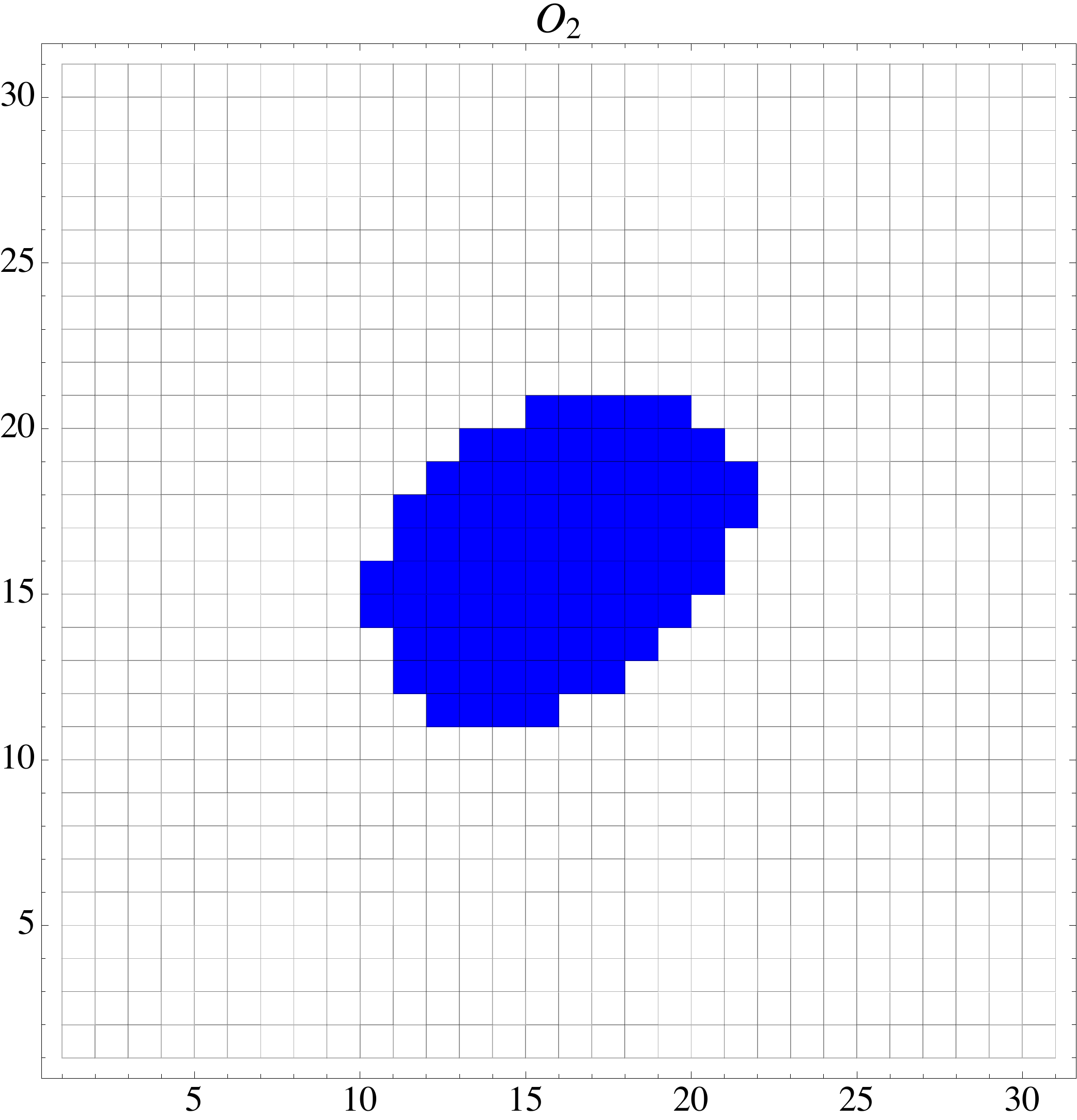}}
\hfill
\subfigure{\includegraphics[width=0.45\textwidth]{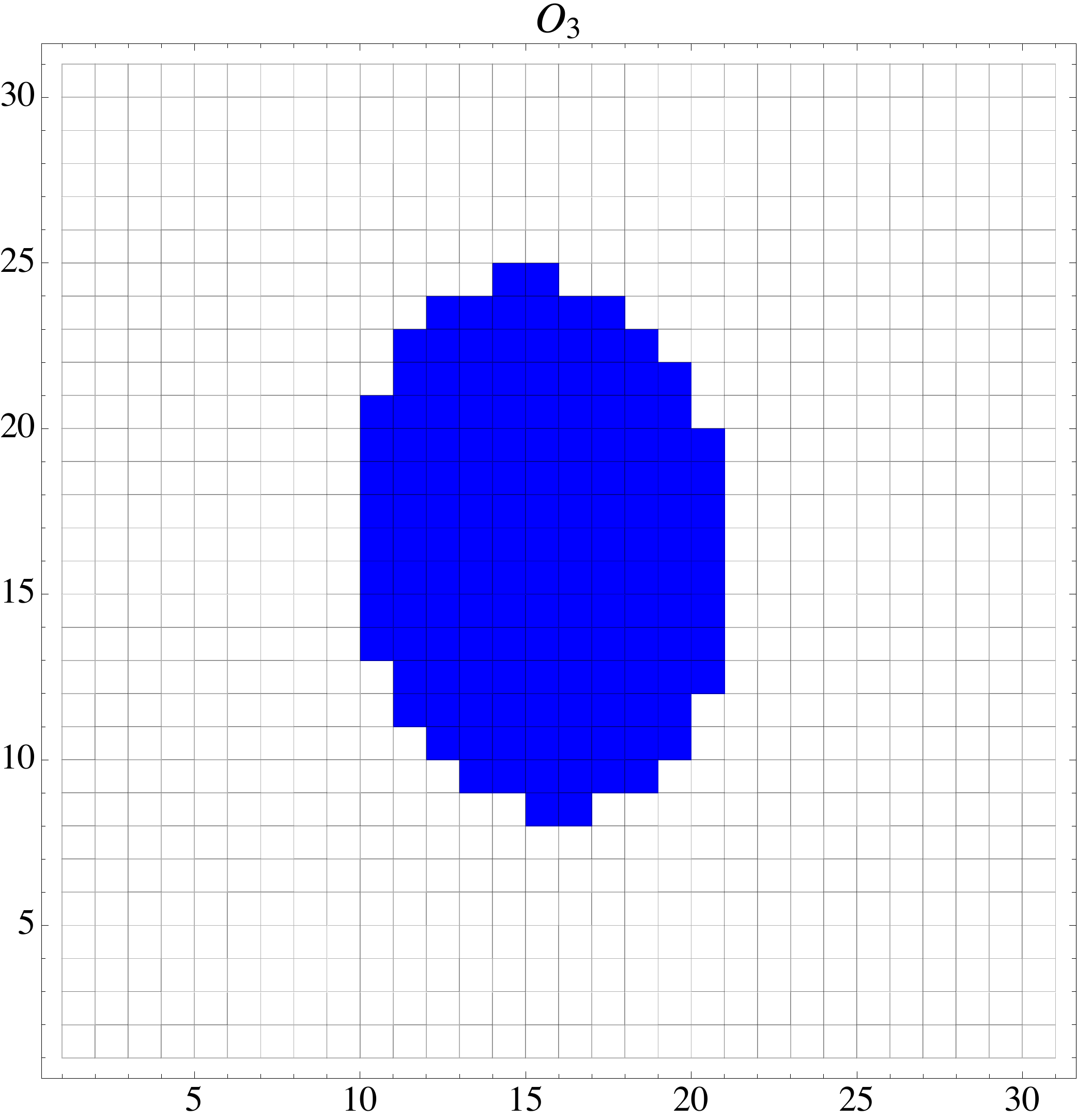}}
\hfill
\caption{The upper left plot shows the final distribution of oscillons, represented by overdensities of the energy density. The upper right and the lower two plots shows the cross sections of different oscillons at a fraction $1/\sqrt{e}$ of their current maximum amplitude. The diameter corresponds to twice the width $R$ as indicated in the upper right figure. Each pixel corresponds to one lattice point. The cross sections are shown at the end of our lattice simulation at $a_{\rm end}=8.3$.} 
\label{fig:eden_field_KKLT}
\end{figure}

On the lattice we find that oscillon formation starts at $a_0\simeq5.8$. At $a=6.41$ we count 4 oscillons. Counting the oscillons at the end of our simulation at $a_{\rm end}=8.3$ is somewhat more ambiguous since some of the regions are very tiny and others could in principle be counted as double or even tripple oscillons. By neglecting the tiny overdensities and counting the spatially large overdensities (double and tripple bubbles) as single oscillons we find $N\simeq 12$ oscillons at $a_{\rm end}=8.3$.
For the semi-analytical calculation of the GW spectrum we assume that the number of oscillons per \textit{physical} volume is constant, i.e.\
\be
N(a) = N(a_0)\cdot\left(\frac{a}{a_0}\right)^3\,,\quad\textrm{with}\quad N(a_0) = 4\,
\en
implying 
\be
N(a_{\rm end}) \simeq 12\,. 
\en

By definition of our oscillon profile eq.~\eqref{eq:oscillon_profile} the width $R$ is defined as the radius of an oscillon at a fraction $1/\sqrt{e}$ of its maximum amplitude.
To estimate the width of the oscillons $R$, we used the spatial field distribution outputted by LATTICEEASY at the end of the lattice simulation and considered the cross sections of different oscillons. The width was then deduced by taking the radius of the contour corresponding to a fraction $1/\sqrt{e}$ of the current amplitude of the oscillon. Some example cross sections are presented in Figure~\ref{fig:eden_field_KKLT}, where each pixel corresponds to one lattice point at $a_{\rm end}=8.3$. The upper right oscillon, for example has a comoving diameter ($\equiv D_c$) of
\be
D_c = 11\,dx = 11\, \frac{L}{512}\,, 
\en
where $dx= L/512$ is the comoving lattice spacing. The physical width is then simply given by
\be
R = 11/2\,dx\,a_{\rm end}\,. 
\en
This can be similarly done for the other two oscillons. By taking the diameter for the other two oscillons (i.e.\ $O_2$ and $O_3$) as indicated in the upper right figure (i.e.\ along the $x$-axis) we obtain the same result for the oscillons $O_2$ and $O_3$.

\subsection{Parameter extraction for hilltop inflation}
\label{sec:parameters_hilltop}
\begin{figure}[h]
\centering
\subfigure{\includegraphics[width=0.45\textwidth]{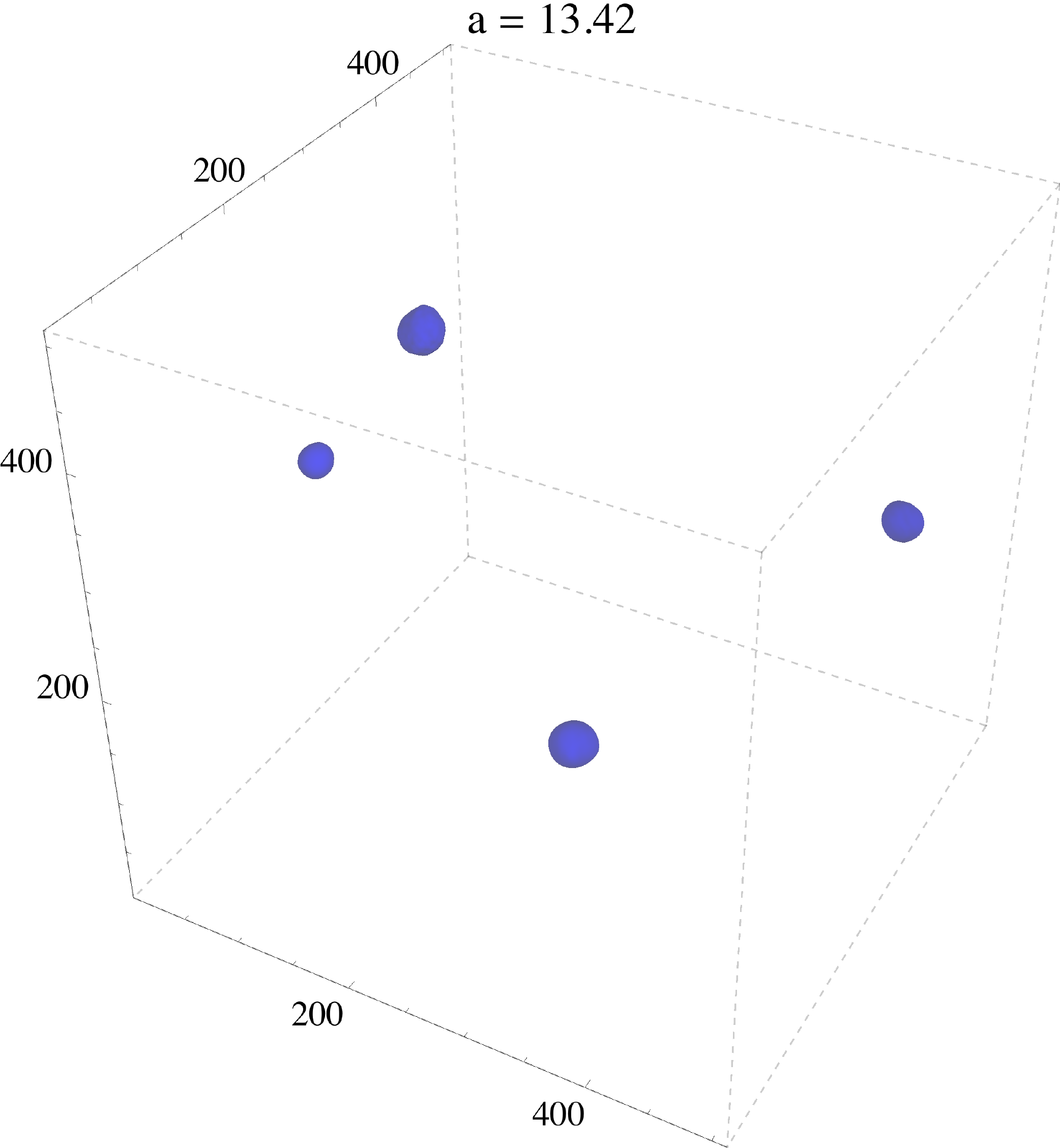}}
\hfill
\subfigure{\includegraphics[width=0.45\textwidth]{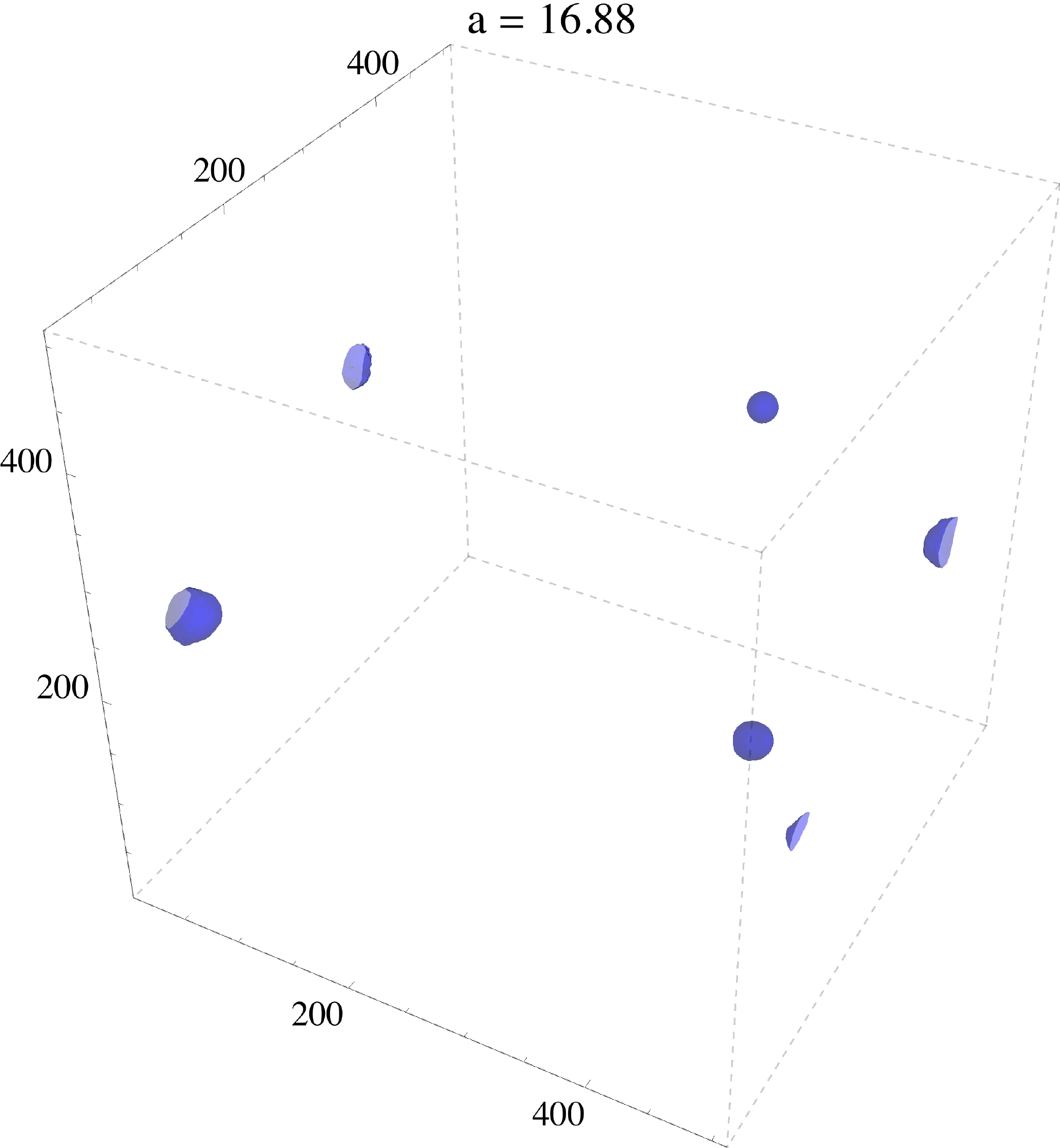}}
\caption{Distribution of the energy density in hilltop inflation represented by regions with $\rho/\langle\rho\rangle \ge 16$ (blue areas). Here we find that the number of oscillons is  $N=4=\textrm{cst.}$ The results were obtained from a lattice simulation with 512 lattice points per spatial dimension.} 
\label{fig:eden_slices_hilltop}
\end{figure}

In the case of hilltop inflation oscillon formation starts at $a_{\rm 0}\simeq5$ in our lattice simulation. Contrary to the KKLT model, where the number of oscillons increases as the universe expands we observe a constant number of oscillons throughout the whole simulation. This renders the task of counting oscillons significantly easier and less ambiguous. 
To determine the number of oscillons we counted again the large energy density overdensities . In Figure~\ref{fig:eden_slices_hilltop} we show the energy density distribution at two different moments in time corresponding to $a=13.42$ and at the end of our lattice simulation $a_{\rm end} = 16.88$. The blue surfaces correspond to regions where $\rho/\langle\rho\rangle \ge 16$. One can see that in both cases the number of oscillons is 
\be
N= 4\,.
\en

\begin{figure}[ht]
\centering
\subfigure{\includegraphics[width=0.45\textwidth]{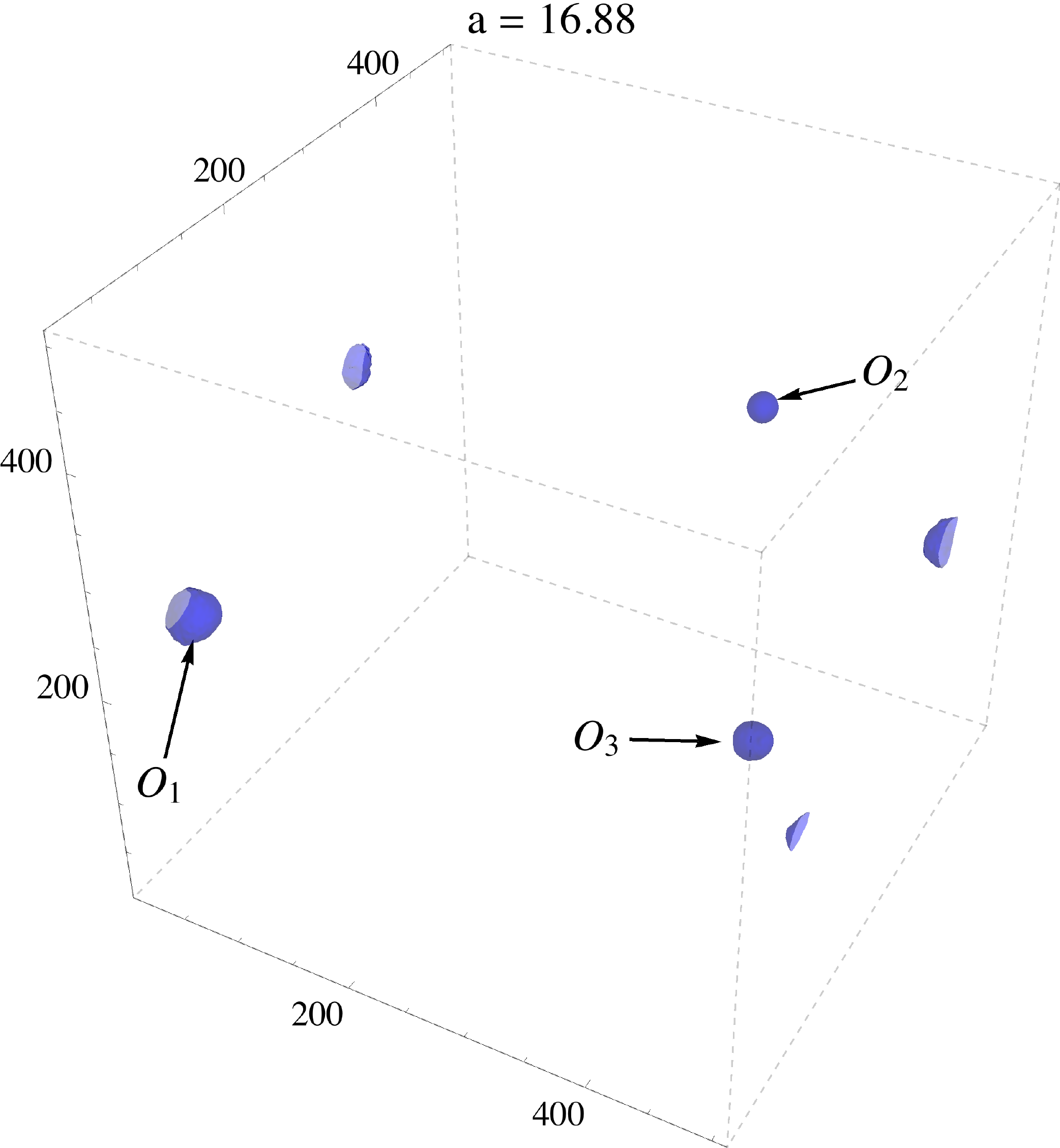}}
\hfill
\subfigure{\includegraphics[width=0.45\textwidth]{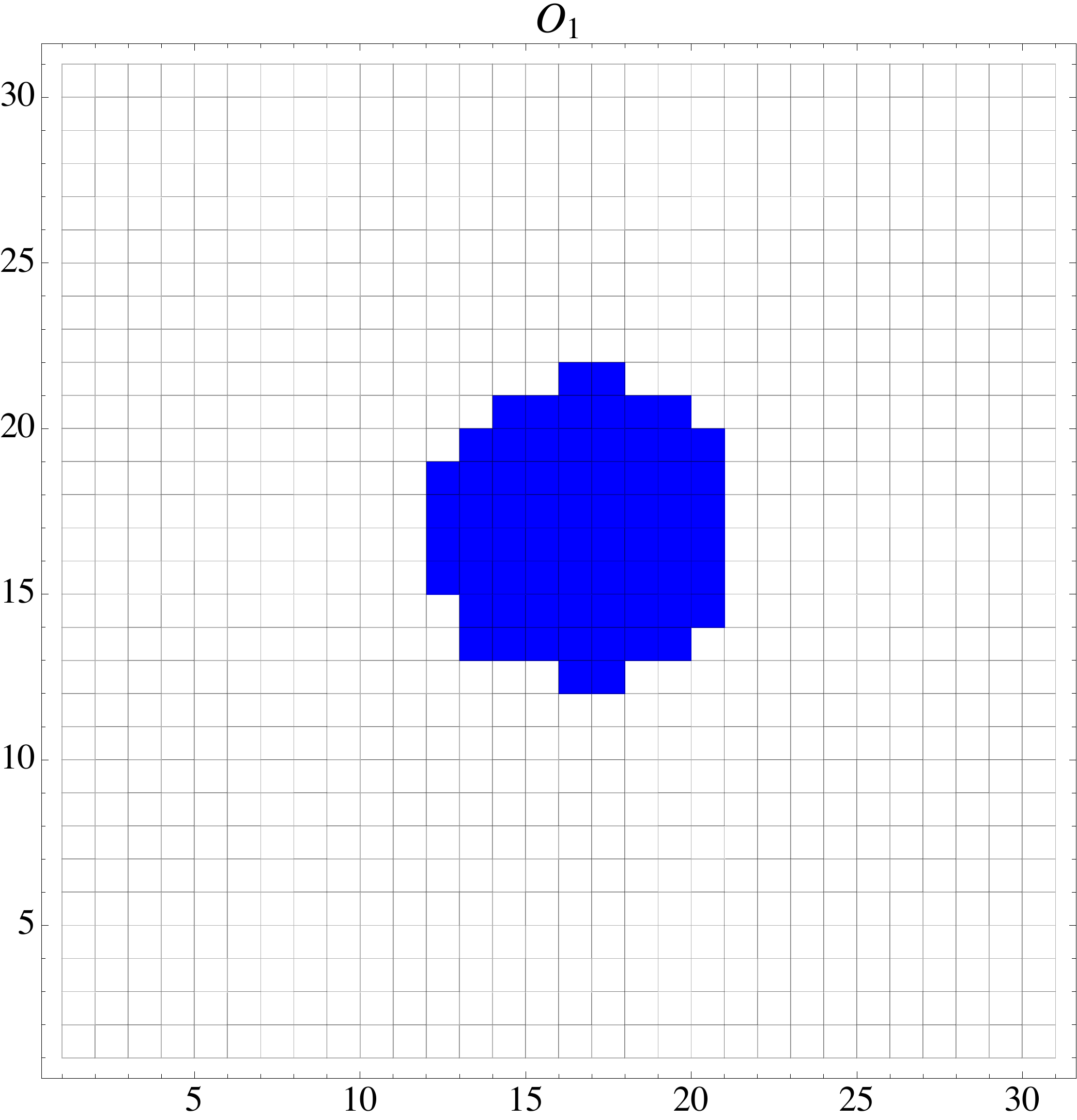}}
\hfill
\subfigure{\includegraphics[width=0.45\textwidth]{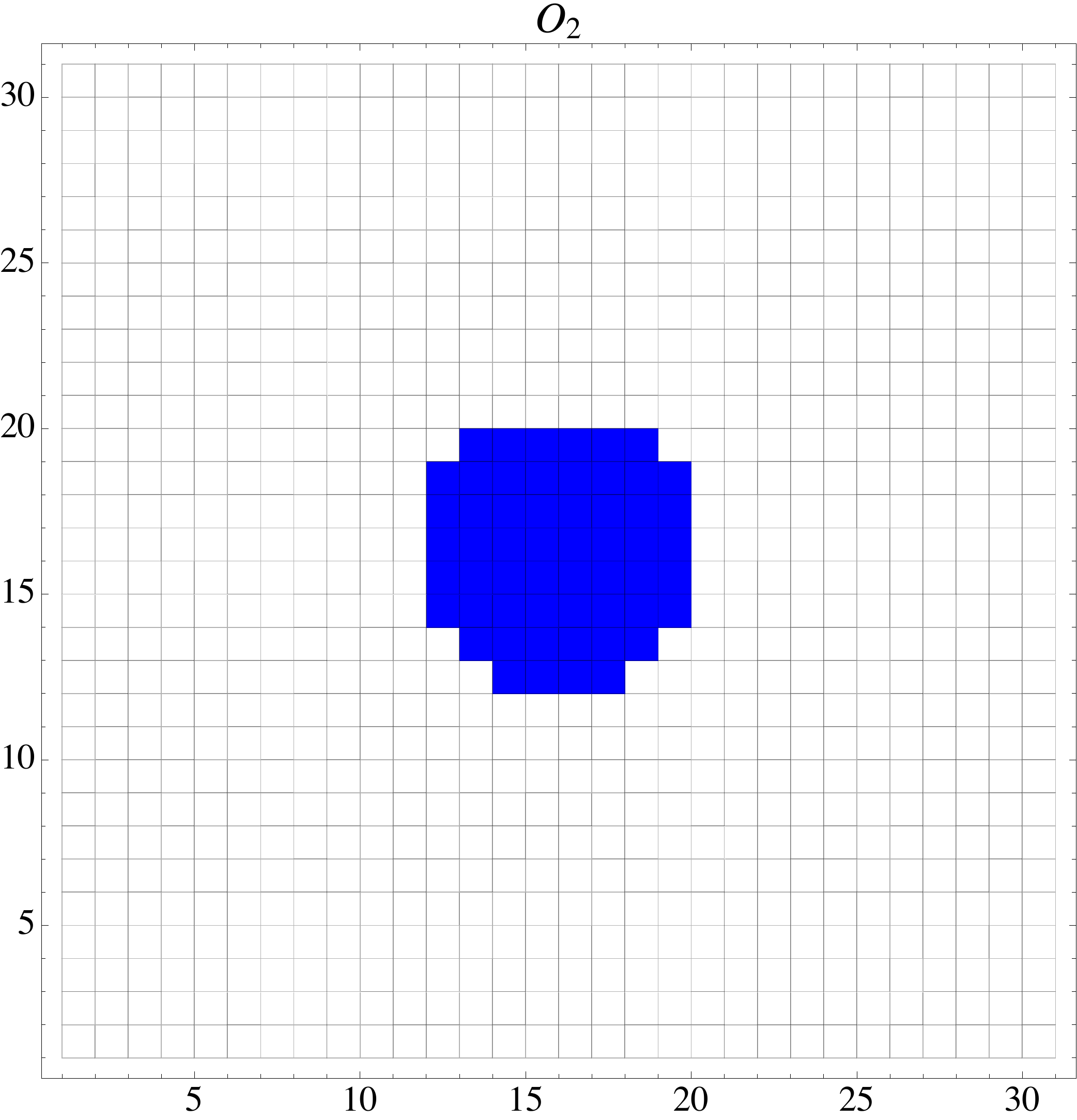}}
\hfill
\subfigure{\includegraphics[width=0.45\textwidth]{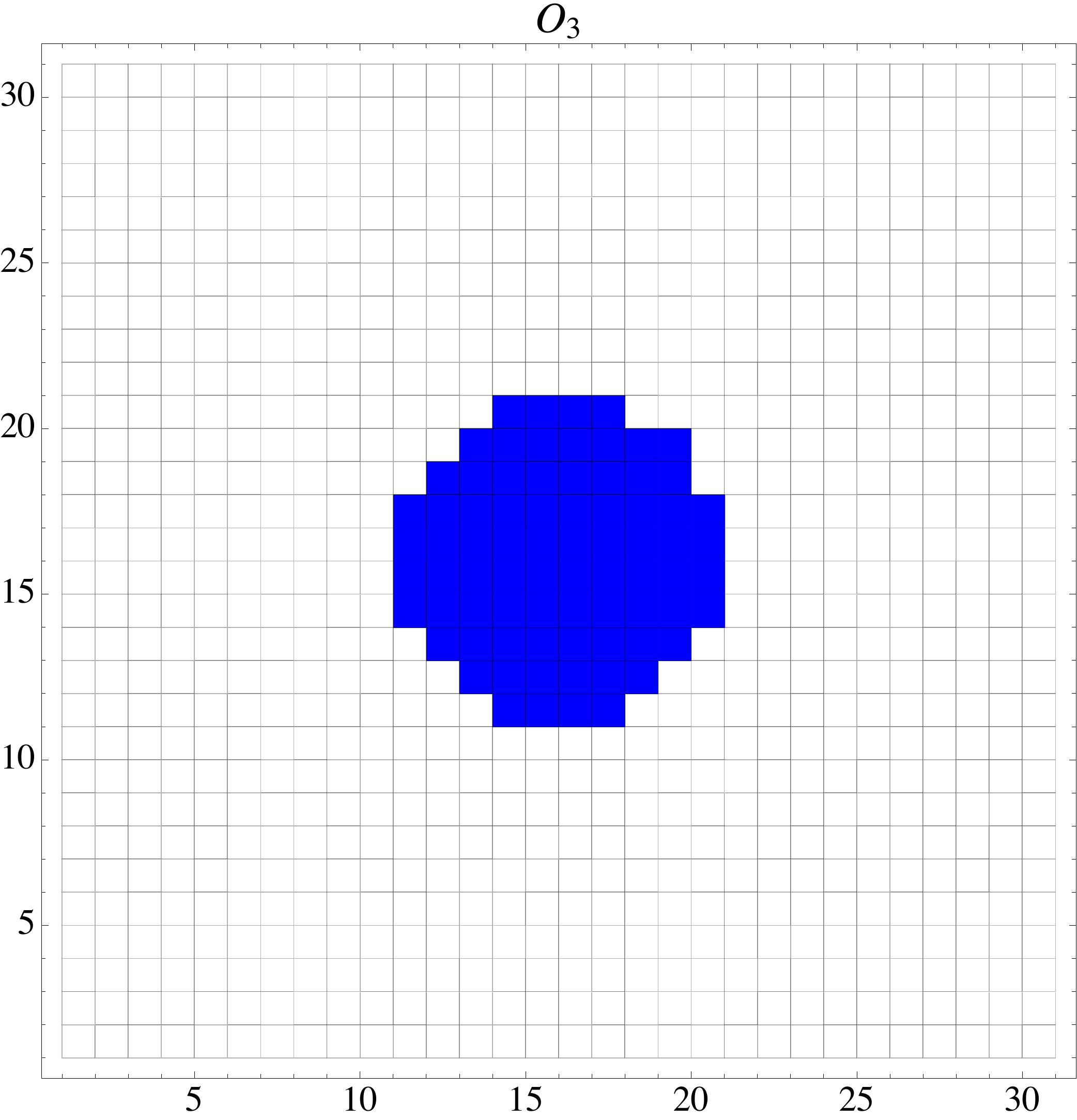}}
\hfill
\caption{Final energy density distribution (upper left) and three example cross sections of different oscillons. The blue contour corresponds to a fraction $1/\sqrt{e}$ of their maximum at the end of the lattice simulation at $a_{\rm end} = 16.88$. Each pixel corresponds to one lattice point. The results were obtained from the lattice simulation of the hilltop inflation model.} 
\label{fig:eden_field_hilltop}
\end{figure}

To extract the width of the oscillons we proceeded in the same way as for the KKLT scenario i.e.\ by taking the width of some example oscillons at a fraction $1/\sqrt{e}$ of their current  amplitude. The examples we used to measure the oscillon width $R$ are shown in Figure~\ref{fig:eden_field_hilltop}. The Figure shows the final distribution of energy density overdensities (upper left) and three example cross sections in field space (upper right and lower plots). The blue contour corresponds to a fraction $1/\sqrt{e}$ of the current oscillon amplitude at the end of our lattice simulation. 
In the case of hilltop inflation we find that the oscillons have a physical width
\be
R = 9/2\,dx\,a_{\rm end}\,. 
\en

\begin{thebibliography}{299}

\bibitem{Ade:2015xua}
  P.~A.~R.~Ade {\it et al.} [Planck Collaboration],
  Astron.\ Astrophys.\  {\bf 594} (2016) A13
  doi:10.1051/0004-6361/201525830
  [arXiv:1502.01589 [astro-ph.CO]].

 
\bibitem{Ade:2015lrj}
  P.~A.~R.~Ade {\it et al.} [Planck Collaboration],
  Astron.\ Astrophys.\  {\bf 594} (2016) A20
  doi:10.1051/0004-6361/201525898
  [arXiv:1502.02114 [astro-ph.CO]].
  
 
\bibitem{Kofman:1994rk}
  L.~Kofman, A.~D.~Linde and A.~A.~Starobinsky,
  Phys.\ Rev.\ Lett.\  {\bf 73} (1994) 3195
  doi:10.1103/PhysRevLett.73.3195
  [hep-th/9405187].
 
\bibitem{Kofman:1997yn}
  L.~Kofman, A.~D.~Linde and A.~A.~Starobinsky,
  Phys.\ Rev.\ D {\bf 56} (1997) 3258
  doi:10.1103/PhysRevD.56.3258
  [hep-ph/9704452].
 
\bibitem{Khlebnikov:1997di}
  S.~Y.~Khlebnikov and I.~I.~Tkachev,
  Phys.\ Rev.\ D {\bf 56} (1997) 653
  doi:10.1103/PhysRevD.56.653
  [hep-ph/9701423].
  
\bibitem{Easther:2006gt}
  R.~Easther and E.~A.~Lim,
  JCAP {\bf 0604} (2006) 010
  doi:10.1088/1475-7516/2006/04/010
  [astro-ph/0601617].
  
\bibitem{Easther:2006vd}
  R.~Easther, J.~T.~Giblin, Jr. and E.~A.~Lim,
  Phys.\ Rev.\ Lett.\  {\bf 99} (2007) 221301
  doi:10.1103/PhysRevLett.99.221301
  [astro-ph/0612294].

\bibitem{GarciaBellido:2007dg}
  J.~Garcia-Bellido and D.~G.~Figueroa,
  Phys.\ Rev.\ Lett.\  {\bf 98} (2007) 061302
  doi:10.1103/PhysRevLett.98.061302
  [astro-ph/0701014].

\bibitem{Dufaux:2007pt}
  J.~F.~Dufaux, A.~Bergman, G.~N.~Felder, L.~Kofman and J.~P.~Uzan,
  Phys.\ Rev.\ D {\bf 76} (2007) 123517
  doi:10.1103/PhysRevD.76.123517
  [arXiv:0707.0875 [astro-ph]].

\bibitem{Zhou:2013tsa}
  S.~Y.~Zhou, E.~J.~Copeland, R.~Easther, H.~Finkel, Z.~G.~Mou and P.~M.~Saffin,
  JHEP {\bf 1310} (2013) 026
  doi:10.1007/JHEP10(2013)026
  [arXiv:1304.6094 [astro-ph.CO]].
  
\bibitem{Figueroa:2013vif}
  D.~G.~Figueroa and T.~Meriniemi,
  JHEP {\bf 1310} (2013) 101
  doi:10.1007/JHEP10(2013)101
  [arXiv:1306.6911 [astro-ph.CO]].


\bibitem{Ashoorioon:2013oha}
  A.~Ashoorioon, B.~Fung, R.~B.~Mann, M.~Oltean and M.~M.~Sheikh-Jabbari,
  JCAP {\bf 1403} (2014) 020
  doi:10.1088/1475-7516/2014/03/020
  [arXiv:1312.2284 [hep-th]].
  
\bibitem{Figueroa:2014aya}
  D.~G.~Figueroa,
  JHEP {\bf 1411} (2014) 145
  doi:10.1007/JHEP11(2014)145
  [arXiv:1402.1345 [astro-ph.CO]].
  
  
\bibitem{Figueroa:2016ojl}
  D.~G.~Figueroa, J.~García-Bellido and F.~Torrentí,
  Phys.\ Rev.\ D {\bf 93} (2016) no.10,  103521
  doi:10.1103/PhysRevD.93.103521
  [arXiv:1602.03085 [astro-ph.CO]].

\bibitem{Antusch:2016con}
  S.~Antusch, F.~Cefala and S.~Orani,
  Phys.\ Rev.\ Lett.\  {\bf 118} (2017) no.1,  011303
  doi:10.1103/PhysRevLett.118.011303
  [arXiv:1607.01314 [astro-ph.CO]].
  
\bibitem{Figueroa:2017vfa}
  D.~G.~Figueroa and F.~Torrenti,
  JCAP {\bf 1710} (2017) no.10,  057
  doi:10.1088/1475-7516/2017/10/057
  [arXiv:1707.04533 [astro-ph.CO]].
  
\bibitem{Liu:2017hua}
  J.~Liu, Z.~K.~Guo, R.~G.~Cai and G.~Shiu,
  arXiv:1707.09841 [astro-ph.CO].
  
\bibitem{Antusch:2017flz}
  S.~Antusch, F.~Cefala, S.~Krippendorf, F.~Muia, S.~Orani and F.~Quevedo,
  arXiv:1708.08922 [hep-th].
 
 
  
\bibitem{Abbott:2007kv}
  B.~P.~Abbott {\it et al.} [LIGO Scientific Collaboration],
  Rept.\ Prog.\ Phys.\  {\bf 72} (2009) 076901
  doi:10.1088/0034-4885/72/7/076901
  [arXiv:0711.3041 [gr-qc]].
 
\bibitem{TheVirgo:2014hva}
  F.~Acernese {\it et al.} [VIRGO Collaboration],
  Class.\ Quant.\ Grav.\  {\bf 32} (2015) no.2,  024001
  doi:10.1088/0264-9381/32/2/024001
  [arXiv:1408.3978 [gr-qc]].
  
\bibitem{AmaroSeoane:2012km}
  P.~Amaro-Seoane {\it et al.},
  GW Notes {\bf 6} (2013) 4
  [arXiv:1201.3621 [astro-ph.CO]].
 
\bibitem{BBO}
  V.~Corbin and N.~J.~Cornish,
  Class.\ Quant.\ Grav.\  {\bf 23} (2006) 2435
  doi:10.1088/0264-9381/23/7/014
  [gr-qc/0512039].
  G.~M.~Harry, P.~Fritschel, D.~A.~Shaddock, W.~Folkner and E.~S.~Phinney,
  Class.\ Quant.\ Grav.\  {\bf 23} (2006) 4887
   Erratum: [Class.\ Quant.\ Grav.\  {\bf 23} (2006) 7361].
  doi:10.1088/0264-9381/23/24/C01, 10.1088/0264-9381/23/15/008
  
\bibitem{Kawamura:2011zz}
  S.~Kawamura {\it et al.},
  Class.\ Quant.\ Grav.\  {\bf 28} (2011) 094011.
  doi:10.1088/0264-9381/28/9/094011
 
 
\bibitem{Punturo:2010zz}
  M.~Punturo {\it et al.},
  Class.\ Quant.\ Grav.\  {\bf 27} (2010) 194002.
  doi:10.1088/0264-9381/27/19/194002
  

\bibitem{GarciaBellido:2007af}
  J.~Garcia-Bellido, D.~G.~Figueroa and A.~Sastre,
  Phys.\ Rev.\ D {\bf 77} (2008) 043517
  doi:10.1103/PhysRevD.77.043517
  [arXiv:0707.0839 [hep-ph]].
 
\bibitem{Dufaux:2010cf}
  J.~F.~Dufaux, D.~G.~Figueroa and J.~Garcia-Bellido,
  Phys.\ Rev.\ D {\bf 82} (2010) 083518
  doi:10.1103/PhysRevD.82.083518
  [arXiv:1006.0217 [astro-ph.CO]].
   
 
\bibitem{Desroche:2005yt}
  M.~Desroche, G.~N.~Felder, J.~M.~Kratochvil and A.~D.~Linde,
  Phys.\ Rev.\ D {\bf 71} (2005) 103516
  doi:10.1103/PhysRevD.71.103516
  [hep-th/0501080].
  
\bibitem{Brax:2010ai}
  P.~Brax, J.~F.~Dufaux and S.~Mariadassou,
  Phys.\ Rev.\ D {\bf 83} (2011) 103510
  doi:10.1103/PhysRevD.83.103510
  [arXiv:1012.4656 [hep-th]].
    
   
\bibitem{Antusch:2015nla}
  S.~Antusch, D.~Nolde and S.~Orani,
  JCAP {\bf 1506} (2015) no.06,  009
  [arXiv:1503.06075 [hep-ph]].
  
  
\bibitem{Amin:2013ika}
  M.~A.~Amin,
  Phys.\ Rev.\ D {\bf 87} (2013) no.12,  123505
  doi:10.1103/PhysRevD.87.123505
  [arXiv:1303.1102 [astro-ph.CO]].

 
 
\bibitem{Copeland:2002ku}
  E.~J.~Copeland, S.~Pascoli and A.~Rajantie,
  Phys.\ Rev.\ D {\bf 65} (2002) 103517
  [hep-ph/0202031].

\bibitem{Broadhead:2005hn}
  M.~Broadhead and J.~McDonald,
  Phys.\ Rev.\ D {\bf 72} (2005) 043519
  [hep-ph/0503081].

\bibitem{Amin:2011hj}
  M.~A.~Amin, R.~Easther, H.~Finkel, R.~Flauger and M.~P.~Hertzberg,
  Phys.\ Rev.\ Lett.\  {\bf 108} (2012) 241302
  [arXiv:1106.3335 [astro-ph.CO]].

\bibitem{Gleiser:2014ipa}
  M.~Gleiser and N.~Graham,
  Phys.\ Rev.\ D {\bf 89} (2014) no.8,  083502
  [arXiv:1401.6225 [astro-ph.CO]].

 
\bibitem{Lozanov:2017hjm}
  K.~D.~Lozanov and M.~A.~Amin,
  arXiv:1710.06851 [astro-ph.CO].
 
\bibitem{Hasegawa:2017iay}
  F.~Hasegawa and J.~P.~Hong,
  arXiv:1710.07487 [astro-ph.CO].

\bibitem{Gleiser:1993pt}
  M.~Gleiser,
  Phys.\ Rev.\ D {\bf 49} (1994) 2978
  [hep-ph/9308279].

\bibitem{Copeland:1995fq}
  E.~J.~Copeland, M.~Gleiser and H.-R.~Muller,
  Phys.\ Rev.\ D {\bf 52} (1995) 1920
  [hep-ph/9503217].
  
\bibitem{Farhi:2005rz}
  E.~Farhi, N.~Graham, V.~Khemani, R.~Markov and R.~Rosales,
  Phys.\ Rev.\ D {\bf 72} (2005) 101701
  [hep-th/0505273].

\bibitem{Fodor:2006zs}
  G.~Fodor, P.~Forgacs, P.~Grandclement and I.~Racz,
  Phys.\ Rev.\ D {\bf 74} (2006) 124003,
  [hep-th/0609023].


\bibitem{Graham:2006vy}
  N.~Graham,
  Phys.\ Rev.\ Lett.\  {\bf 98} (2007) 101801
  [hep-th/0610267].

\bibitem{Gleiser:2007te}
  M.~Gleiser and J.~Thorarinson,
  Phys.\ Rev.\ D {\bf 76} (2007) 041701
  [hep-th/0701294 [HEP-TH]].

\bibitem{Achilleos:2013zpa}
  V.~Achilleos, F.~K.~Diakonos, D.~J.~Frantzeskakis, G.~C.~Katsimiga, X.~N.~Maintas, E.~Manousakis, C.~E.~Tsagkarakis and A.~Tsapalis,
  Phys.\ Rev.\ D {\bf 88} (2013) 045015
  [arXiv:1306.3868 [hep-th]].

\bibitem{Bond:2015zfa}
  J.~R.~Bond, J.~Braden and L.~Mersini-Houghton,
  JCAP {\bf 1509} (2015) no.09,  004
  [arXiv:1505.02162 [astro-ph.CO]].


\bibitem{Kachru:2003aw}
  S.~Kachru, R.~Kallosh, A.~D.~Linde and S.~P.~Trivedi,
  Phys.\ Rev.\ D {\bf 68} (2003) 046005
  doi:10.1103/PhysRevD.68.046005
  [hep-th/0301240].
  
\bibitem{Balasubramanian:2005zx}
  V.~Balasubramanian, P.~Berglund, J.~P.~Conlon and F.~Quevedo,
  JHEP {\bf 0503} (2005) 007
  doi:10.1088/1126-6708/2005/03/007
  [hep-th/0502058].
  
\bibitem{Conlon:2005ki}
  J.~P.~Conlon, F.~Quevedo and K.~Suruliz,
  JHEP {\bf 0508} (2005) 007
  doi:10.1088/1126-6708/2005/08/007
  [hep-th/0505076].

\bibitem{Silverstein:2008sg}
  E.~Silverstein and A.~Westphal,
  Phys.\ Rev.\ D {\bf 78} (2008) 106003
  doi:10.1103/PhysRevD.78.106003
  [arXiv:0803.3085 [hep-th]].
  
\bibitem{McAllister:2008hb}
  L.~McAllister, E.~Silverstein and A.~Westphal,
  Phys.\ Rev.\ D {\bf 82} (2010) 046003
  doi:10.1103/PhysRevD.82.046003
  [arXiv:0808.0706 [hep-th]].


\bibitem{Graham:2006xs}
  N.~Graham and N.~Stamatopoulos,
  Phys.\ Lett.\ B {\bf 639} (2006) 541
  doi:10.1016/j.physletb.2006.06.070
  [hep-th/0604134].
  
\bibitem{Gleiser:2009ys}
  M.~Gleiser and D.~Sicilia,
  Phys.\ Rev.\ D {\bf 80} (2009) 125037
  doi:10.1103/PhysRevD.80.125037
  [arXiv:0910.5922 [hep-th]].
  
\bibitem{Amin:2010jq}
  M.~A.~Amin and D.~Shirokoff,
  Phys.\ Rev.\ D {\bf 81} (2010) 085045
  doi:10.1103/PhysRevD.81.085045
  [arXiv:1002.3380 [astro-ph.CO]].
  
  
\bibitem{Saffin:2014yka}
  P.~M.~Saffin, P.~Tognarelli and A.~Tranberg,
  JHEP {\bf 1408} (2014) 125
  doi:10.1007/JHEP08(2014)125
  [arXiv:1401.6168 [hep-ph]].


  
\bibitem{Felder:2000hq}
  G.~N.~Felder and I.~Tkachev,
  Comput.\ Phys.\ Commun.\  {\bf 178} (2008) 929
  doi:10.1016/j.cpc.2008.02.009
  [hep-ph/0011159].
  
  
\bibitem{Antusch:2014qqa}
  S.~Antusch, D.~Nolde and S.~Orani,
  JCAP {\bf 1405} (2014) 034
  doi:10.1088/1475-7516/2014/05/034
  [arXiv:1402.5328 [hep-ph]].

\bibitem{Antusch:2015vna}
  S.~Antusch, F.~Cefala, D.~Nolde and S.~Orani,
  JCAP {\bf 1602} (2016) no.02,  044
  doi:10.1088/1475-7516/2016/02/044
  [arXiv:1510.04856 [hep-ph]].
  
\bibitem{Antusch:2015ziz}
  S.~Antusch and S.~Orani,
  JCAP {\bf 1603} (2016) no.03,  026
  doi:10.1088/1475-7516/2016/03/026
  [arXiv:1511.02336 [hep-ph]].
  
\bibitem{Polarski:1995jg}
  D.~Polarski and A.~A.~Starobinsky,
  Class.\ Quant.\ Grav.\  {\bf 13} (1996) 377
  doi:10.1088/0264-9381/13/3/006
  [gr-qc/9504030].
  
\bibitem{Khlebnikov:1996mc}
  S.~Y.~Khlebnikov and I.~I.~Tkachev,
  Phys.\ Rev.\ Lett.\  {\bf 77} (1996) 219
  doi:10.1103/PhysRevLett.77.219
  [hep-ph/9603378].
  
\end{thebibliography}
\end{document}